\documentclass[11pt]{article}

\usepackage{amsmath,amssymb}
\usepackage{graphicx}

\makeatletter
 
  \@addtoreset{equation}{section}
 \makeatother
\def\href#1#2{#2}
\def \II {I\hspace{-.1em}I\hspace{.1em}}
\def \IIA {\mbox{\II A\hspace{.2em}}}

\newcommand{\bea}{\begin{eqnarray}}
\newcommand{\eea}{\end{eqnarray}}
\newcommand{\be}{\begin{equation}}
\newcommand{\ee}{\end{equation}}
\newcommand{\nn}{\nonumber}
\newcommand{\pa}{\partial}

\newcommand{\la}{\langle}
\newcommand{\ra}{\rangle}

\renewcommand{\a}{\alpha}
\renewcommand{\b}{\beta}
\newcommand{\e}{\epsilon}
\newcommand{\ve}{\varepsilon}
\newcommand{\s}{\sigma}

\newcommand{\tr}{{\rm tr}}
\newcommand{\hPhi}{\hat{\Phi}}

\newcommand{\hcH}{\hat{{\cal H}}}
\newcommand{\hLambda}{\hat{\Lambda}}

\newcommand{\hvarphi}{\hat{\varphi}}

\newcommand{\hx}{\hat{x}}
\newcommand{\hpa}{\stackrel{\circ}{\partial}}

\newcommand{\Ord}{{\cal O}}

\newcommand{\A}{(\mbox{\bf Ad}[A])}

\begin{document}

\renewcommand{\thefootnote}{\arabic{footnote}}
\begin{titlepage}

\begin{center}

\hfill 
\vskip 1.0in

{\bf \LARGE More on the Nambu-Poisson M5-brane Theory}\\[5mm]
{\sl \Large Scaling limit, background independence and}\\[2mm]
{\sl \Large an all order solution to the Seiberg-Witten map}

\vskip .7in
{\sc 
Chien-Ho CHEN$,^{a,}$\footnote{chenchienho@gmail.com} 
Kazuyuki FURUUCHI$,^{b,}$\footnote{furuuchi@phys.cts.nthu.edu.tw}
Pei-Ming HO$\,{}^{a,}$\footnote{pmho@phys.ntu.edu.tw}\\[1mm]
and Tomohisa TAKIMI$\,{}^{a,}$\footnote{tomo.takimi@gmail.com}}\\
\vskip 10mm
{\sl
${}^a$Department of Physics and Center for Theoretical Sciences, \\
National Taiwan University, 
Taipei 10617, Taiwan,
R.O.C.}\\
\vspace{3mm}
{\sl
${}^b$National Center for Theoretical Sciences\\
National Tsing-Hua University, Hsinchu 30013, Taiwan, R.O.C.}\\
\noindent{ \smallskip }\\

\vspace{1pt}
\end{center}
\begin{abstract}
We continue our investigation on the Nambu-Poisson 
description of M5-brane 
in a large constant $C$-field background
(NP M5-brane theory)
constructed in Refs.\cite{Ho:2008nn,Ho:2008ve}.
In this paper, the low energy limit where the NP M5-brane theory is applicable is clarified.
The background independence of the NP M5-brane theory
is made manifest using the variables in the BLG model of multiple M2-branes.
An all order solution to the Seiberg-Witten map is also constructed.
\end{abstract}

\end{titlepage}

\setcounter{footnote}{0}

\section{Introduction}


M2-branes and M5-branes are the fundamental building blocks of M-theory.
As a way to understand the mysterious nature of M-theory,
it is desirable to understand them as much as possible.
However, for the time being our understanding of the M-branes are 
far less comprehensive than our understanding of D-branes in string theory, 
and comparatively the M5-brane is even less understood than the M2-brane.

The solitonic solution for M5-branes \cite{Gueven:1992hh}
in 11 dimensional supergravity was found even before 
the advent of M-theory \cite{Witten:1995ex,Hull:1994ys}.
After that,  
people successfully constructed 
the equations of motion \cite{Howe:1996yn,Howe:1997fb}
and then the action for a single M5-brane 
\cite{Pasti:1997gx,Bandos:1997ui,Aganagic:1997zq,Bandos:1997gm}. 
The quantum aspects of a single M5-brane 
were also understood to some extent \cite{Witten:1996hc,Belov:2006jd}.
However, while we understand the physics of multiple D-branes
and non-commutative D-branes, 
the analogous knowledge about M5-branes is absent.
In this paper,
we make an effort to understand
a related problem---
the physics of M5-brane in $C$-field background.

Recently, a new worldvolume action 
describing a single M5-brane
in a large constant $C$-field background 
was constructed \cite{Ho:2008nn,Ho:2008ve}.
This action was obtained from 
the BLG model of multiple M2-branes 
\cite{Bagger:2006sk,Bagger:2007jr,Gustavsson:2007vu},
which has a gauge symmetry based on Lie 3-algebra.
By choosing the Nambu-Poisson (NP) structure as the Lie 3-algebra
and expanding around a certain background, one obtains
the new M5-brane action.
We will refer to this theory as {\it NP M5-brane theory} in the following.
The construction is analogous to that of 
a D(p+2)-brane in a constant $B$-field background 
from infinitely many Dp-branes, and in fact
it can be uplifted to M-theory in certain cases
through the relation between M-theory and type \IIA superstring theory.
Since extensive research has been made
on D-branes in a constant $B$-field background (see e.g. \cite{Douglas:2001ba}
and references therein),
the uplift to M-theory will give us good clues to understand 
the M5-brane worldvolume theory. 
Indeed, taking the
analogy with D-branes in a constant $B$-field background as a guidance,
in Refs.\cite{Ho:2008nn,Ho:2008ve} 
it was conjectured that 
the NP M5-brane theory is related to the 
conventional description of M5-brane 
\cite{Howe:1996yn,Howe:1997fb,Pasti:1997gx,Bandos:1997ui,Aganagic:1997zq,Bandos:1997gm}
in a constant $C$-field background
through the so-called Seiberg-Witten map \cite{Seiberg:1999vs}.
In the case of D-branes in a constant $B$-field background
one has two descriptions of the same system,
the one using ordinary coordinates and another with 
non-commutative coordinates.
Seiberg-Witten map relates
the ordinary description and the non-commutative description of 
D-branes in a constant $B$-field background.
Therefore, this conjecture is a natural extension of
such a D-brane system to 
an M5-brane in a constant $C$-field background.
In Ref.\cite{Ho:2008ve}, the Seiberg-Witten map for M5-brane
was constructed
up to the first order in 
a parameter which 
parametrizes the strength of the interaction through the NP bracket.
We remind the reader that
in the case of D-branes in a constant $B$-field background,
non-commutative description was practically much more convenient
than the ordinary description in the zero-slope limit \cite{Seiberg:1999vs}, 
and the same will be true
for the M5-brane in a constant $C$-field background.
In this sense, we may say that the NP bracket description
captures the structure of the M5-brane worldvolume theory
in this background in a more essential way.

Several non-trivial supports for this conjecture 
have been given in the original papers as well as
in subsequent works.
In the original papers \cite{Ho:2008nn,Ho:2008ve},
it was shown that the NP M5-brane theory has 
the same field contents with conventional 
M5-branes 
as well as
the six-dimensional (2,0) supersymmetry.
It was also shown that
the double dimensional reduction of the NP M5-brane theory
reduces to the Poisson description of D4-brane 
in a constant $B$-field background in a rather non-trivial way.
This, through the M-theory -- \IIA string relation,
provides an indirect support for 
the identification of NP M5-brane theory as a theory of M-theory five-brane.
Furthermore,
an argument within M-theory based on 
the central charge of the 
eleven-dimensional super-Poincare algebra
(or ``M-theory superalgebra") 
in the BLG model was given in \cite{Furuuchi:2008ki}.
In \cite{Furuuchi:2009zx},
the BPS string solitons
on the NP M5-brane worldvolume was constructed
and compared with the corresponding object
in the ordinary description \cite{Howe:1997ue,Michishita:2000hu,Youm:2000kr}
via the Seiberg-Witten map,
and precise match was found up to the first order
in the NP parameter for the scalar field configurations.
The test was further extended to the comparison
of defining BPS equations for string solitons 
between two descriptions
in \cite{Furuuchi:2010sp}. 
Notice that these are also direct tests of the conjecture
without referring to the M-theory -- \IIA string relation.
And it has been clarified
how the self-dual relations,
which is a salient feature of the M5-brane theory
\cite{Howe:1996yn,Howe:1997fb,Pasti:1997gx,Howe:1997vn,Schwarz:1997mc},
are encoded in the NP M5-brane action
\cite{Ho:2008ve,Pasti:2009xc,Furuuchi:2010sp}.
With those evidences in hand,
now the conjectured equivalence between two descriptions
of M5-brane worldvolume theory in a constant $C$-field background
has become very plausible.
This 
in turn would provide a support 
for the validity of the BLG model as a description of M-theory branes.

In this paper, we will collect further evidences
for the conjectured equivalence between
the NP M5-brane theory and 
conventional M5-brane theory in a constant $C$-field background.
In section \ref{secPre},
we recall some background materials
which are useful for later sections.
In section \ref{secNBM5lim},
we identify the low energy limit
where the NP M5-brane theory is applicable.
In section \ref{secBGind},
the background independence of the NP M5-brane theory
is made manifest by using the background independent variables in the BLG model.
We also identify ``open membrane metric" which governs
the propagation of fields in the NP M5-brane worldvolume,
and effective tension of the NP M5-brane.
In section \ref{secSWmap},
we construct an all order solution to the Seiberg-Witten map.
We end this paper with the discussions on future directions.

\section{Preliminaries}\label{secPre}

In this section, we recall
some background materials
which are useful in later sections.

\subsection{The BLG model of multiple M2-branes}\label{BLG}

The M5-brane action of Refs.\cite{Ho:2008nn,Ho:2008ve}
was constructed from 
the Bagger-Lambert-Gustavsson model (BLG model)
of multiple M2-branes 
\cite{Bagger:2006sk,Bagger:2007jr,
Gustavsson:2007vu}.
This model has a novel type of gauge symmetry 
based on an algebraic structure 
called Lie 3-algebra \cite{Filippov}.
For a linear space 
${\cal V} = \{ \sum_{a=1}^{\dim {\cal V}} v_a T^a; v_a \in \mathbb{C} \}$,
Lie 3-algebra structure is defined by a tri-linear map
which is called 3-bracket
$[*,*,*]$ : ${\cal V}^{\otimes 3} \rightarrow {\cal V}$,
satisfying the following properties:\\
\ \\ 
1. Skew-symmetry:
\be
 \label{skew}
[A_{\s(1)}, A_{\s(2)} , A_{\s(3)}] = (-1)^{|\s|} [A_1, A_2, A_3].
\ee
2. Fundamental identity:
\bea 
 \label{FI} 
&&[A_1, A_2, [B_1, B_2, B_3]] \nn \\
&=& [[A_1,A_2,B_1],B_2,B_3] + [B_1,[A_1,A_2,B_2],B_3] + [B_1,B_2,[A_1,A_2,B_3]].\nn\\
\eea
A linear space endowed with a Lie 3-algebra structure
will be called Lie 3-algebra.
In terms of the basis $T^a$, Lie 3-algebra can be expressed in terms of
the structure constants $f^{abc}{}_d$:
\bea
 \label{st}
[T^a,T^b,T^c] = f^{abc}{}_d T^d .
\eea
We will be interested in Lie 3-algebra with inner product
$\la *, *\ra$ 
${\cal V} \otimes {\cal V} \rightarrow {\mathbb{C}}$ (metric Lie 3-algebra):
\bea
\la T^a,T^b \ra = h^{ab} ,
\eea
so that we can construct an action.
We refer to $h^{ab}$ as metric of the Lie 3-algebra. 
%
We require following invariance of the inner product
which
is needed for the gauge invariance of the BLG model:
\bea
 \label{invm}
\la [T^a, T^b, T^c], T^d \ra 
+ \la T^c, [T^a, T^b, T^d] \ra = 0.
\eea
%
Together with the skew-symmetry property (\ref{skew}),
the invariance of the metric
(\ref{invm}) requires the indices of structure constants
$f^{abcd} \equiv f^{abc}{}_{e} h^{ed}$
to be totally anti-symmetric: 
\bea
 \label{tantis}
f^{abcd} = 
\frac{1}{4!}f^{[abcd]}  .
\eea
The action of the BLG model is given by
\bea
S = \int d^3 x \; {\cal L}, 
\eea
where the Lagrangian density ${\cal L}$ is given by
\bea
\label{BLaction}
&&{\cal L} = -\frac{1}{2} \la D^{\mu}\phi^I, D_{\mu} \phi^I\ra 
+ \frac{i}{2} \la\bar\Psi, \Gamma^{\mu}D_{\mu}\Psi\ra 
+\frac{i}{4} \la\bar\Psi, \Gamma_{IJ} [\phi^I, \phi^J, \Psi]\ra \nn \\ 
&&\qquad \quad -V(\phi) + {\cal L}_{CS}. 
\eea
$\phi^I = \phi^I_a T^a$ ($I = 1,\cdots,8$)
are scalar fields on the worldvolume
which describe embedding of the M2-brane worldvolume in the transverse
eight dimensions in the eleven-dimensional target space-time.
$\Psi = \Psi_a T^a$
are Majorana spinors on 1+2 dimensional worldvolume,
but can be combined into a single Majorana spinor in eleven dimensions
subject to the chirality condition 
$\Gamma \Psi = - \Psi$, $\Gamma \equiv \Gamma_{012}$.
$D_{\mu}$ is the covariant derivative 
\be
 \label{Dmu}
(D_\mu  \varphi(x))_a = \partial _{\mu} \varphi_a(x) -\tilde{A}_\mu{}^b{}_a(x) \varphi_b(x), 
\quad 
\tilde{A}_{\mu}{}^b{}_a \equiv A_{\mu cd} f^{cdb}{}_a ,
\ee
where $A_\mu$ is the gauge field
and $\varphi$ collectively represents $\phi^I$ and $\Psi$.
$V(\phi)$ is the potential 
\bea
 \label{BLGpot}
V(\phi) = \frac{1}{12}\la [\phi^I, \phi^J, \phi^K], [\phi^I, \phi^J, \phi^K]\ra .
\eea
The Chern-Simons term for the gauge potential is given by
\bea
\label{CS}
{\cal L}_{CS} = \frac{1}{2}\ve^{\mu\nu\lambda}
\left(f^{abcd}A_{\mu ab}\pa_{\nu}A_{\lambda cd} 
+ \frac{2}{3} f^{cda}{}_g f^{efgb} A_{\mu ab} A_{\nu cd} A_{\lambda ef} \right). 
\eea
The action is invariant under
the following gauge transformation:
\bea
 \label{gauge}
\delta_\Lambda \phi^I_a &=& 
\Lambda_{cd}[T^c,T^d,\phi^I]_a =\Lambda_{cd} f^{cde}{}_a \phi^I_e
=\tilde{\Lambda}^e{}_a  \phi^I_e, \nn\\
\delta_\Lambda \Psi_a &=&  
\Lambda_{cd}[T^c,T^d,\Psi]_a 
= \Lambda_{cd} f^{cde}{}_a \Psi_e
= \tilde{\Lambda}^e{}_a  \Psi_e, \nn\\
\delta_\Lambda \tilde{A}_{\mu}{}^b{}_a 
&=&
\pa_\mu  \tilde{\Lambda}_{\mu}{}^b{}_a 
-\tilde{\Lambda}^b{}_{c} \tilde{A}_{\mu}{}^c{}_a
+ \tilde{A}_{\mu}{}^b{}_c  \tilde{\Lambda}^c{}_{a},
\quad \tilde{\Lambda}^b{}_{a} 
\equiv f^{cdb}{}_a \Lambda_{cd}. 
\eea

\subsection{M-theory -- type \IIA superstring relation}\label{secMIIA}

M-theory and type \IIA superstring theory are related
by a circle compactification of M-theory.
We will
study an M5-brane in a constant $C$-field background,
which is an uplift of a D4-brane in a constant $B$-field background.
Although in the case of D-branes
in a constant $B$-field background
information was extracted from the worldsheet theory
of open string,
extracting information from
the quantization of 
M2-brane worldvolume theory
in a constant $C$-field background
is more complicated
(see \cite{Kawamoto:2000zt,Ho:2007vk,Hofman:2002rv,Hofman:2002jz,Chu:2009iv} 
for some approaches from the M2-brane worldvolume).
Instead, we will make use
of the M-theory -- \IIA string relation to uplift the results
in the D-branes in a constant $B$-field background.
We briefly review the M-theory -- \IIA string relation in this subsection.

Let us consider a compactification of M-theory
on a circle
with 
the coordinate compactification radius $R_{10}^{coord}$
(here we compactify the $x^{10}$ direction).
We take the coordinates where
the ten-dimensional part of the background metric
of M-theory and that of type \IIA string theory are the same:
\bea
g_{\mu\nu\, (M)} = g_{\mu\nu\, (IIA)} \equiv g_{\mu\nu}, \quad \mbox{for } \mu,\nu = 0,\cdots 9.
\eea
The relation of the M-theory parameters and those in type \IIA superstring theory 
are given as
\bea
 \label{MIIA}
R_{10}^{phys} = g_s \ell_s, \quad \ell_P = g_s^{1/3} \ell_s ,
\eea
where 
$R_{10}^{phys}$ is the physical compactification
radius measured by the M-theory metric $g_{10,10\,(M)}$
\bea
(R_{10}^{phys})^2 \equiv g_{10,10\,(M)}(R_{10}^{coord})^2 ,
\eea
and $\ell_s \equiv (\a')^{1/2}$ and
$\ell_P$ is the eleven-dimensional Planck scale
(we follow the convention in Polchinski's text book \cite{Polchinski:1998rr})
which is related to the M-theory brane tensions as
\bea
 \label{TM25}
T_{M2} = \frac{1}{(2\pi)^2 \ell_P^{3}},
\quad
T_{M5} = \frac{1}{2\pi} (T_{M2})^2 = \frac{1}{(2\pi)^5 \ell_P^6}.
\eea
By the circle compactification of M-theory,
M2-branes which wrap on the circle become fundamental strings,
and those
which do not wrap on the circle become D2-branes.
Similarly,
M5-branes which wrap on the circle become D4-branes,
and those which do not wrap on the circle become NS5-branes.
The M-theory -- \IIA string relation (\ref{MIIA}) correctly reproduces the
tensions
of D2-brane, D4-brane, fundamental string and NS5-brane
which are given by
\bea
T_{Dp} &=& \frac{1}{(2\pi)^p g_s \ell_s^{p+1}} ,\\
T_{F1} &=& \frac{1}{2\pi\a'} , \\
T_{NS5} &=& \frac{1}{(2\pi)^5 g_s^2 \ell_s^6} .
\eea

\subsection{Open string theory in a constant $B$-field background}\label{openB}

Open string theory on D-branes in a constant $B$-field background
can be described by gauge theory on non-commutative space
\cite{Chu:1998qz,Chu:1999gi,Schomerus:1999ug,Seiberg:1999vs}.
Many interesting results have been obtained,
such as Seiberg-Witten map \cite{Seiberg:1999vs}, 
non-commutative instantons/solitons 
\cite{Nekrasov:1998ss,Furuuchi:1999kv,Furuuchi:2000vx,Ho:2000ea,Gopakumar:2000zd}
and UV-IR mixing \cite{Minwalla:1999px}.
Some of these results should have corresponding
uplift in M-theory via the M-theory -- \IIA string relation
discussed in the previous section,
which we would like to investigate.
Let us briefly review some results in 
open strings in a constant $B$-field background.

In a constant $B$-field background,
the propagation of open strings 
is governed by the
so-called open string metric,
and the effective coupling constant
is also modified, which is often called 
open string coupling.
Those are given as \cite{Seiberg:1999vs}
\bea
 \label{OpenFreedom}
&&
\left( \frac{1}{G+2\pi\a'\Phi} + \frac{\theta}{2\pi\a'} \right)^{ij}
= 
\left(\frac{1}{g + 2\pi\a'B}\right)^{ij} , \nn \\
&&G_s =
g_s \left(\frac{\det (G+ 2\pi\a'\Phi)}{\det (g + 2\pi\a'B)} \right)^{1/2} ,
\eea
where $G_{ij}$ is the open string metric and
$G_s$ is the open string coupling.
$\Phi$ parametrizes a freedom in the description \cite{Seiberg:1999vs}.
A natural choice for $\Phi$ is
\bea
 \label{choicePhi}
  \Phi = -B,
\eea
which leads to
\bea 
G^{ij} &=& - \frac{1}{(2\pi\a')^2} \left( \frac{1}{B} g \frac{1}{B} \right)^{ij}, \label{OSinvmet} \\
G_{ij} &=& - (2\pi\a')^2 \left( B g^{-1} B \right)_{ij}, \label{OSmet} \\
\theta^{ij} &=& \left( \frac{1}{B} \right)^{ij} , \label{theta} \\
G_s &=& g_s \det (2\pi\a'Bg^{-1})^{1/2}. \label{OSC}
\eea

In general, 
even if we restrict ourselves to the massless sector,
the low energy effective field theory on D-branes
in such background still receives $\a'$ corrections
and is described by an action
like Nambu-Goto-Dirac-Born-Infeld type action
on non-commutative space, or with further $\a'$ corrections.
On the other hand, the non-commutative Yang-Mills theory (NCYM)\footnote{%
By Yang-Mills theory we refer to the theory
described by the action with the curvature square term 
$\mbox{tr} F^2$. 
We include the case where the gauge group is $U(1)$ for brevity,
since on the non-commutative space the action for 
the case with $U(1)$ gauge group takes the similar
form to that of the case with $U(N)$ gauge group
due to the self-coupling of the gauge fields through the non-commutativity.}
is obtained in a particular zero-slope limit
\cite{Seiberg:1999vs}:
\bea
 \label{FTlim}
\a' &\sim& \e^{1/2} \rightarrow 0 , \nn \\
g_{ij} &\sim& \e \rightarrow 0 .
\eea
Notice that this limit 
with finite $B$ leads to the finite open string metric.
The Yang-Mills coupling on the
D$p$-brane is given by
\bea
 \label{YMC}
\frac{1}{g_{YM}^2}
=
\frac{(\a')^{\frac{3-p}{2}}}{(2\pi)^{p-2}G_s}
=
\frac{(\a')^{\frac{3-p}{2}}}{(2\pi)^{p-2}g_s}
\left( \frac{\det (g + 2\pi\a'B) }{\det G} \right)^{\frac{1}{2}} .
\eea
From (\ref{YMC}) it follows that 
to obtain a finite Yang-Mills coupling
in the zero-slope limit (\ref{FTlim}), 
we should scale $g_s$ and $G_s$ as
\bea
 \label{scl}
G_s &\sim& \e^{\frac{3-p}{4}}, \nn \\
g_s &\sim& \e^{\frac{3-p+r}{4}} ,
\eea
where $r$ is the rank of the background $B$-field.

\section{The NP M5-brane theory limit from the zero-slope limit of 
type \IIA string theory in a constant $B$-field background}\label{secNBM5lim}

We would like to study the situation
where
the NP M5-brane theory 
reduces to the Yang-Mills theory on a Poisson manifold
as the D4-brane worldvolume theory
upon double dimensional reduction. 
Here, the description on a Poisson manifold can be regarded either
as a small non-commutativity approximation of 
the Moyal product description 
(explained in section \ref{secBGind}),
or
another description of the D4-brane in a constant $B$-field background.
In the former case, 
using the M-theory -- type \IIA superstring relation
reviewed in the previous subsection,
we can translate  
the scaling to the non-commutative Yang-Mills description in 
type \IIA superstring discussed in subsection \ref{openB} to
the scaling limit which leads to the NP M5-brane theory.\footnote{%
We assume that we are working in a particular choice of the freedom
in the descriptions which might be there in the M5-brane theory,
as in (\ref{OpenFreedom}) in the case of D-branes in a constant $B$-field background.}
However, one should be aware of the difficulty
in obtaining the non-commutative Yang-Mills description of D4-brane
from the double dimensional reduction of the
deformation of NP M5-brane theory
\cite{Chen:2010ny}.
On the other hand,
when we take the latter interpretation,
we will assume that the open string metric and open string coupling
are the same both 
in the non-commutative description and
the Poisson description of the D4-brane in a $B$-field background.

Now let us consider the double dimensional reduction 
of the NP M5-brane action.
Here, we study the configuration
where the worldvolume of the NP M5-brane extends in (012345)-directions,
among which (012) were the worldvolume directions 
of the original multiple M2-branes.
Unlike sec.~\ref{secMIIA},
in this section 
we compactify the $x^5$-direction instead of the $x^{10}$-direction.   
Then, the M-\IIA relation (\ref{MIIA}) together with
the zero-slope limit (\ref{FTlim})
enforce the following scaling of the parameters in M-theory:
\bea
\ell_P &\sim& \e^{1/3}, \label{AFTlimlp}  \\
R_{5}^{phys} &\sim& \e^{1/2} ,\label{AFTlimL5} 
\eea
where $R_{5}^{phys}$ is the physical compactification
radius in the $5$-th direction.
Eq. (\ref{AFTlimlp}) ensures the decoupling of the eleven-dimensional gravity
(more discussions on this point later.)
Eq. (\ref{AFTlimL5}) means that if we fix 
(i.e., do not scale with $\e$)
the coordinate compactification length,
the $g_{55}$ component of the metric 
scales as $g_{55} \sim \e$ in the zero-slope limit (\ref{FTlim}).
Notice that this behavior is the same as 
the scaling of $g_{33}$ and $g_{44}$ in 
(\ref{FTlim}).
We take 
the coordinate compactification radius as $R_{5}^{coord}$.
It is related to the physical compactification
radius $R_5^{phys}$ as follows:
\bea
(R_5^{phys})^2 = g_{55} (R_{5}^{coord})^2 .
\eea
The $C$-field in M-theory is related to the $B$-field in \IIA string theory
as
\bea
 \label{CB}
C_{345} (2\pi R_{5}^{coord}) = B_{34}. 
\eea
Summarizing, we can define the NP M5-brane theory limit
by 
\bea
 \label{NBM5lim}
\ell_P &\sim& \e^{1/3} , \nn \\
g_{ij(M)} &\sim& \e , \nn \\
C_{ijk} &\sim& \e^0 \quad (i,j = 3,4,5).
\eea
Notice that although the scaling limit (\ref{NBM5lim}) 
was extracted from the scaling limit 
(\ref{FTlim})
in \IIA string theory via the M-theory - \IIA string relation,
the limit itself 
can be taken without the compactification in the $x^5$ direction.
In other words, the limit (\ref{NBM5lim}) can be studied
totally within M-theory.

If we further tune the scaling in (\ref{NBM5lim}) 
so that the effective tension of M2-branes becomes
finite, we arrive at the so-called OM-theory \cite{Gopakumar:2000ep}.
However, since we are interested in the field theory description
of M5-brane, we will not consider the limit to the OM-theory.

The scaling of the $C_{012}$ component
of the background $C$-field is not independently chosen
from the NP M5-brane limit (\ref{NBM5lim}),
since
it must obey the
non-linear self-dual relations \cite{Howe:1996yn,Howe:1997fb}:
\bea
 \label{SDC}
\frac{\sqrt{-\det g}}{6} \e_{\mu_1\mu_2\mu_3\mu_4\mu_5\mu_6}
C^{\mu_4\mu_5\mu_6}
&=&
\frac{1+K}{2} g_{\mu_1 \mu}(\tilde{G}^{-1})^{\mu\nu} C_{\nu \mu_2 \mu_3}, \nn\\
&&(\mu_1,\cdots,\mu_6, \mu, \nu = 0,\cdots,5),
\eea
where $\e_{\mu_1\mu_2\mu_3\mu_4\mu_5\mu_6}$ is a totally anti-symmetric tensor
with $\e_{012345} = 1$,
and
\bea
 \label{KtG}
K = \sqrt{1+\frac{1}{24} (2\pi)^4 \ell_P^6 C^2},
\quad
\tilde{G}_{\mu\nu}
=
\frac{1+K}{2K} \left( g_{\mu\nu}+\frac{1}{4} (2\pi)^4 \ell_P^6 C^2_{\mu\nu} \right),
\eea
\bea
C^2 \equiv C_{\mu_1\mu_2\mu_3} C_{\nu_1\nu_2\nu_3} g^{\mu_1\nu_1}g^{\mu_2\nu_2}g^{\mu_3\nu_3},
\quad
(C^2)_{\mu \nu} \equiv 
C_{\mu \mu_2\mu_3} C_{\nu \nu_2\nu_3} g^{\mu_2\nu_2}g^{\mu_3\nu_3}.
\eea
Using (\ref{SDC}), one can check that
$C_{012} \sim \e^{-1}$ 
with the finite metric $g_{\mu\nu}=\eta_{\mu\nu}$ (for $\mu,\nu=0,1,2$),
using the non-linear self-dual relations for ordinary M5-brane.
On the other hand,
if we use the coordinates 
where $g_{ij}=\eta_{ij}$ $(i,j =3,4,5)$,
the NP M5-brane limit
(\ref{NBM5lim})
amounts to 
take $C_{345} \sim \e^{-3/2}$
(note that $C_{\mu\nu\rho}$ is a tensor and 
the value of the components depend on the coordinate system).
Comparing with this, 
one excepts that
$C_{012}$ is not strong enough to induce 
finite interaction
through the NP bracket in the (012)-directions 
in the scaling limit (\ref{NBM5lim})
(as long as we do not tune $C_{012}$ to reach to the OM-theory).
We give more explicit arguments in appendix \ref{AppScaling}.

The NP M5-brane action 
is an analogue of the Poisson bracket Yang-Mills action
on D4-brane, and indeed 
it 
reduces
to it 
upon double dimensional reduction albeit rather non-trivially \cite{Ho:2008ve}
(notice that the self-dual two-form gauge field 
on the M5-brane reduces to the one-form gauge field
on D4-brane without the self-dual relations).
The use of the Poisson bracket Yang-Mills action
is justified in the particular scaling limit
(\ref{FTlim}) from which we obtained
the scaling limit (\ref{NBM5lim}).
Therefore, the scaling limit (\ref{NBM5lim}) 
is also required to justify the use of the 
NP M5-brane action. 

One would like to describe the 
M5-brane theory obtained through
the scaling limit (\ref{NBM5lim})
by quantities which remain finite in this limit,
analogous to the open string metric (\ref{OSmet})
and the non-commutative parameter (\ref{theta})
in the case of open string theory in a constant $B$-field background.
This will be achieved in section \ref{secBGind}.

\section{Background independence of NP M5-brane theory
and open membrane metric}\label{secBGind}

In \cite{Seiberg:2000zk},
it was shown that
when we
obtain NCYM 
as an expansion around a background 
in the matrix model,
the background independence 
becomes manifest.
Similar story holds
when we construct 
NP M5-brane action from an expansion around
a background in the BLG model.

At the time when 
uplift of open string theory in a constant $B$-field background
to M-theory was studied,
the ``open membrane metric" as
the M-theory analogue of open string metric
\cite{Bergshoeff:2000jn,Bergshoeff:2000ai,Gibbons:2000ck,%
VanderSchaar:2001ay,Berman:2001rka,Bergshoeff:2001xx}
has been proposed.
We make
an observation that
the open membrane metric
(in the scaling limit (\ref{NBM5lim}))
appears rather naturally 
in the kinetic term of the embedding coordinate fields
in our construction.
The effective tension of the NP M5-brane
can also be read off from the kinetic term
of the embedding coordinate fields.

\subsection{Manifest background independence of NCYM 
on a D4-brane from D2-branes}\label{BGID2}

Let us first recall the
background independence of NCYM discussed in
\cite{Seiberg:1999vs,Seiberg:2000zk}.
The background independence here means that
we hold closed string variables $g_s$ and $g_{ij}$ fixed
when we vary
the non-commutative parameter $\theta^{ij}$. 

For our purpose of comparing the NP M5-brane action with
the NCYM 
action on D4-brane 
with the non-commutativity in (34)-directions,
it is convenient to start 
from the action for multiple D2-branes.
The potential term in the
low energy effective action on D2-branes is given by
\bea
 \label{D2pot}
\frac{1}{(2\pi)^2 g_s \ell_s^3}
\int d^3 x \,
\frac{1}{(2\pi\a')^2}
\frac{1}{4}
g_{II'}g_{JJ'}\,
\tr [X^I,X^J][X^{I'},X^{J'}],\quad (I,J = 3,\cdots,9),
\eea
where $X^I$'s are Hermitian matrices with mass dimension $[X] = -1$. 
Here and throughout this paper,
we will use $[A]$ to express the mass dimension of a quantity $A$.

Let us consider the background $X^{i}_{bg} = \hat{x}^i$ ($i=3,4$) satisfying
\bea
 \label{NCBG}
[\hx^i ,\hx^j ] = i \theta^{ij}  ,
\eea
where $\theta^{ij}$ is an anti-symmetric constant tensor 
with mass dimension $[\theta^{ij}] = -2$.
The algebra (\ref{NCBG}) can be realized
by matrices with infinite size,
which is interpreted as infinitely many D2-branes.
We parametrize the fluctuation around the background (\ref{NCBG}) as
\bea
X^i = \hx^i + \theta^{ij} \hat{A}_j (\hx).
\eea
The mass dimension of $\hat{A}_i$ is 
$[\hat{A}_i]=1$, which is the standard mass dimension
when the Yang-Mills coupling is an overall factor of the Yang-Mills action.
To discuss the background independence,
it is convenient to introduce variables $C_i$ as
\bea
 \label{Ci}
C_i \equiv B_{ij} X^j = B_{ij} \hx^j + \hat{A}_i,
\eea
where $B_{ij} = (\theta^{-1})_{ij}$, presuming
the relation (\ref{theta}).

The covariant derivatives in NCYM 
can be written using $C_i$ as
\bea
D_i \hvarphi 
= \partial_i \hvarphi - i [\hat{A}_i, \hvarphi] 
= -i [C_i, \hvarphi].
\eea
It follows that
\bea
 \label{F-B}
-i[C_i,C_j]= \hat{F}_{ij} - B_{ij} .
\eea
The open string metric and open string coupling
are given as in (\ref{OSmet}), (\ref{OSC}):
\bea
G_{ij} &=& - (2\pi\a')^2 (B g^{-1} B )_{ij} , \\
G_s &=& g_s \det (2\pi\a' B g^{-1})^{1/2} 
= g_s \sqrt{ \det G}\, \frac{|\mbox{Pf}\, \theta|}{2\pi\a'}.
\eea

On the other hand, 
the algebra of trace-class
infinite size matrices can be 
isomorphically mapped
to the algebra of
square-integrable functions on ${\mathbb R}^2$,
with the product given by the so-called Moyal (star) product
(see e.g. \cite{Aoki:1999vr}).
If we denote the matrices by $\hat{\ }$ on the symbols, 
the map is given as
\bea
\hat{f}(\hat{x}) = \int d^2k\, f(k) e^{ik\hat{x}}
&\leftrightarrow & 
f(x) = \int d^2k\, f(k) e^{ikx} , \label{map} \\
\hat{f} \hat{h} &\leftrightarrow & f(x) \ast h (x) , \label{star} \\
\tr &\leftrightarrow & \int \frac{d^{2} x}{2\pi} \frac{1}{|\mbox{Pf}\, \theta|}  , \label{trace}
\eea
where $f(x)$ and $h(x)$ are square integrable functions on ${\mathbb R}^2$, 
and $\ast$ is the Moyal product:
\bea
 \label{Moyal}
f(x) \ast h(x) \equiv
e^{\frac{i}{2}\theta^{ij} \frac{\pa}{\pa z^i} \frac{\pa}{\pa x^j} }
f(z) h(x) \Bigr|_{z=x}  .
\eea
Notice that 
in the lowest order in the expansion in $\theta^{ij}$,
the anti-symmetrized Moyal products
reduces to the Poisson bracket:
\bea
 \label{PB}
f(x) \ast h(x) - h(x) \ast f(x) 
&=& i \theta^{ij} \frac{\pa}{\pa x^i} f(x) \frac{\pa}{\pa x^j} h(x) + \Ord (\theta^2) \nn \\
&=& i \{ f(x) , h(x) \}_{Poisson} + \Ord (\theta^2),
\eea
where the Poisson bracket is given by
\bea
 \label{Poisson}
\{ f(x) , h(x) \}_{Poisson} =
\theta^{ij} \frac{\pa}{\pa x^i} f(x) \frac{\pa}{\pa x^j} h(x) .
\eea

Using (\ref{map})--(\ref{trace}), we obtain
\bea
 \label{BGIaction}
&&\frac{1}{(2\pi)^3 g_s \ell_s^{3}} 
\int d^{3}x \,
\frac{1}{(2\pi\a')^2}
\frac{1}{4}
g_{II'}g_{JJ'} \tr [X^I,X^J][X^{I'},X^{J'}] 
\quad (I,J = 3,\cdots , 9)\nn \\
&=&
\frac{1}{(2\pi)^{5} G_s \ell_s^{5}} \int d^{5} x \,
\sqrt{\det G} \, 
\Biggl[
-(2\pi\a')^2
\frac{1}{4}
G^{ii'} G^{jj'} (\hat{F}_{ij} - B_{ij})(\hat{F}_{i'j'} - B_{i'j'}) \nn\\
&&\qquad + 
\frac{1}{2} g_{II'}G^{ii'} D_i X^I D_{i'} X^{I'} 
+ \frac{1}{(2\pi\a')^2} \frac{1}{4} g_{II'}g_{JJ'} [X^I,X^J][X^{I'},X^{J'}]
\Biggr] \nn \\
&=&
\frac{1}{g_{YM}^2} \int d^{5} x \,
\sqrt{\det G} \, 
\Biggl[
-\frac{1}{4} G^{ii'} G^{jj'} (\hat{F}_{ij} - B_{ij})(\hat{F}_{i'j'} - B_{i'j'}) \nn\\
&&\qquad + \frac{1}{2} g_{II'}G^{ii'} D_i \phi^I D_{i'} \phi^{I'} 
+ \frac{1}{4} g_{II'}g_{JJ'} [\phi^I,\phi^J][\phi^{I'},\phi^{J'}]
\Biggr]  ,
 \nn \\
&&\qquad \quad (i,j = 3,4; \, I,J = 5,\cdots, 9),
\eea
where
\bea
\frac{1}{g_{YM}^2} 
\equiv \frac{(2\pi\a')^2}{(2\pi)^{5} G_s \ell_s^{5}}
= \frac{1}{(2\pi)^3 G_s \ell_s} ,
\eea
and
\bea
\phi^I \equiv \frac{1}{2\pi\a'} X^I .
\eea
In the above, 
with a slight abuse of notation,
we have identified matrices and functions on ${\mathbb R}^2 $
through the map (\ref{map}) and used the same symbols.
(For example, $\hat{F}_{ij}$ above should be read as function on ${\mathbb R}^2 $ 
which is mapped from the matrix defined in (\ref{F-B}) with the same symbol
through the map (\ref{map})).
As mentioned before, 
the background independence here means
we hold the closed string variables $g_s$ and $g_{ij}$ fixed,
and the change in the non-commutative parameter $\theta^{ij}$ 
arises only from the 
change of the background (\ref{NCBG}).
In the first line of (\ref{BGIaction}),
the background independence is manifest
since the change in the non-commutative parameter $\theta^{ij}$
is totally due to the choice of the background in (\ref{NCBG})
and the closed string metric and closed string coupling are fixed.
Notice that not only on the left hand side but
also on the right hand side of (\ref{trace})
the background independence of the measure is also clear,
since the rescaling of the non-commutative parameter can be 
generated by the rescaling of coordinates $x^i$,
which cancel with each other in the measure (\ref{trace}).

To discuss background independence without taking the zero-slope limit,
we can consider a more general action, see \cite{Seiberg:2000zk}.
On the other hand, 
we are interested 
in the zero-slope limit
where the closed string metric $g_{ij}$ is scaled as in (\ref{FTlim}).

\subsection{Manifest background independence of 
NP M5-brane action from the BLG model}\label{BGINP5}

We would like to proceed in a parallel way 
when constructing the NP M5-brane action from the BLG model.
To extract the essential point,
we will focus on the potential term of the BLG model (\ref{BLGpot}):
\bea
 \label{pot}
\int d^3 x \, V(\phi)
=
\int d^3 x \,
\frac{1}{12}
g_{II'}g_{JJ'}g_{KK'} \,
\la [\phi^I,\phi^J,\phi^K] , [\phi^{I'},\phi^{J'},\phi^{K'}] \ra , 
\eea
where $\phi^I$ is a canonically normalized scalar field in three dimensions,
i.e. $[\phi^I]=1/2$.
We have also introduced the target space metric $g_{IJ}$
in (\ref{pot}) in order to take into account the scaling limit (\ref{NBM5lim}).
We will refer to the target space metric
$g_{IJ}$ as ``closed membrane metric,"
taking an analogy with the closed string metric in the 
case of open string theory in a constant $B$-field background
explained in subsection \ref{openB}.

To obtain the target space interpretation,
we define
\bea
 \label{TrgtX}
X^I \equiv \phi^I \left((2\pi\right)^{2/3}\ell_P)^{3/2} .
\eea
Then, $X^I$ has a dimension of length, $[X^I] = -1$.
The potential term (\ref{pot}) takes the form
\bea
 \label{potX}
 \int d^3 x \, V(X)
&=&
\frac{1}{(2\pi)^{2}\ell_P^3}
\int d^3 x \,
\frac{1}{(2\pi)^{4}\ell_P^6}
\frac{1}{12}
g_{II'}g_{JJ'}g_{KK'}\,
\la [X^I,X^J,X^K] , [X^{I'},X^{J'},X^{K'}] \ra , \nn \\
&& \qquad (I,J = 3,\cdots,10) .
\eea

Next, we choose 
the Nambu-Poisson structure on ${\mathbb R}^3$ 
as the Lie 3-algebra structure of the BLG model
(see subsection \ref{BLG})
\cite{Ho:2008nn,Ho:2008ve}.
The Lie 3-bracket is given by the NP bracket
\bea
 \label{Our3bra}
[A,B,C] = \{ A,B,C\} =
\theta^{ijk} \frac{\pa}{\pa x^i} A(x) \frac{\pa}{\pa x^j} B(x)  \frac{\pa}{\pa x^k} C(x),
\quad (i,j = 3,4,5) .
\eea
Here, we choose $\theta^{ijk}$ to be a constant totally anti-symmetric tensor.
The mass dimension of the Nambu-Poisson tensor $\theta^{ijk}$ is $[\theta^{ijk}]=-3$.
Notice that the NP bracket is a natural generalization
of the Poisson bracket (\ref{Poisson}).

The elements of the Lie 3-algebra are given
by square-integrable functions on ${\mathbb R}^3$.\footnote{%
The background needs not be square-integrable,
and indeed the M5-brane background 
(\ref{M5bg}) which we will discuss shortly
is an example of such background.
See \cite{Furuuchi:2008ki} for further discussions on this point,
and see \cite{Ho:2009nk} for an alternative
description of such background by introducing
non-positive definite metric of the Lie 3-algebra.}
The inner product of the Lie 3-algebra is given by
\bea
 \label{tr3}
\la A, B \ra = \int \frac{d^3x}{2\pi}  \frac{1}{|\theta^{345}|}  \, A(x) B(x). 
\eea
The normalization of the inner product
(\ref{tr3}) is chosen in a parallel way to that in (\ref{trace}) in the case of matrix model
(the choice of the $2\pi$ factor is for convenience
in the comparison with NCYM upon double dimensional reduction).
This choice ensures the {background independence}.

The M5-brane extending in (012345)-direction is obtained by 
expanding 
the multiple M2-brane action extending in (012)-directions
around the background
\bea
 \label{M5bg}
X^i_{bg} = x^i, \quad (i=3,4,5).
\eea
We parametrize the fluctuation around the background as
\bea
X^i = x^i + \hat{b}^i(x) = x^i + \frac{1}{2} \theta^{ijk} \hat{b}_{jk}(x).
\eea
The mass dimension of the field $\hat{b}_{ij}$ is $[\hat{b}_{ij}]=2$.
The field strength of the two-form gauge potential
in $(345)$-directions is defined as
\bea
\qquad
\hcH_{ijk}
=
C_{ijk} \,
\left(
\frac{1}{6}
C_{\ell m n}
 \left( 
\{X^\ell,X^m,X^n \} 
- \theta^{\ell mn}
 \right)
\right) ,
\eea
where 
we have defined the totally anti-symmetric tensor $C_{ijk}$ by 
\bea
 \label{Cth}
- C_{ijk} \theta^{k\ell m} = \delta_i^{\ell} \delta_j^m - \delta_i^m \delta_j^{\ell}.
\eea
We postulate that the
tensor $C_{ijk}$ is a component of the background $C$-field.
We also postulate that the ``(inverse) open membrane metric" is given by
\bea
 \label{GOM}
G^{ii'}_{OM} = 
\frac{1}{2}
\frac{\theta^{ijk}}{(2\pi)^2\ell_P^3} \frac{\theta^{i'j'k'}}{(2\pi)^2\ell_P^3} g_{jj'} g_{kk'}
\quad (i,j,k = 3,4,5) .
\eea
Not like in the case of string theory in a constant $B$-field background
where we can read off the open string metric and the
non-commutative parameter from the worldsheet two-point function,
here
we do not have a derivation
of our postulates from the M2-brane worldvolume theory.
We will show that our postulates are consistent 
with the open string metric and the open string coupling
after the double dimensional reduction of the NP M5-brane action.
This reasoning was basically the same as the one used in \cite{VanderSchaar:2001ay},
though our study will be
restricted to the 
scaling limit (\ref{NBM5lim}).

We define the covariant derivatives in (345)-directions as
\bea
D_i \hat{\varphi} 
\equiv
- \frac{1}{2}
C_{ijk} \{X^j,X^k, \hat{\varphi} \} \quad (i,j = 3,4,5).
\eea
Now the potential term (\ref{potX}) is rewritten as
\bea
 \label{NBM5H2}
&&\int d^3x\, V(X) \nn \\
&=&
T_6
\int d^6x \, \sqrt{\det G_{OM}}\, 
\Biggl[ 
\frac{1}{T_6} \frac{1}{2\pi} \frac{1}{12} G^{ii'}_{OM} G^{jj'}_{OM} G^{kk'}_{OM} 
(\hcH_{ijk} - C_{ijk})
(\hcH_{i'j'k'} - C_{i'j'k'}) \nn \\
&&+ 
\frac{1}{2}
g_{II'} G^{ii'}_{OM} D_{i} X^I D_{i'} X^{I'}
+ \frac{1}{(2\pi)^4 \ell_P^6} 
\frac{1}{4} g_{ii'}g_{JJ'}g_{KK'}\, \la [X^i,X^J,X^K], [X^{i'},X^{J'},X^{K'}] \ra \nn \\
&&+ \frac{1}{(2\pi)^4 \ell_P^6} 
\frac{1}{12} g_{II'}g_{JJ'}g_{KK'}\, \la [X^I,X^J,X^K], [X^{I'},X^{J'},X^{K'}] \ra
 \Biggr]\nn \\
&& \qquad (i,j = 3,4,5; \, I,J = 6,\cdots, 10),
\eea
where 
\bea
 \label{T6}
T_6
\equiv
\frac{T_{M2}}{ 2\pi |\theta^{345}| \sqrt{\det G_{OM}}}
\eea
is the effective tension of the NP M5-brane 
which is read off from the kinetic term for $X^I$.
Notice that the effective tension of the NP M5-brane
is much smaller than the fundamental M5-brane tension 
$T_{M5}$ when $\theta \ll \ell_P^3$.
Thus the back reaction to the closed membrane metric
is negligible in the limit $\ell_P \rightarrow 0$.
In the above, by $\det G_{OM}$ we mean the determinant of $(G_{OM})_{ii'}$
which is the inverse of $G_{OM}^{ii'}$ defined in (\ref{GOM}).
The rewriting of the
gauge field kinetic term is not as
obvious 
as in the case of non-commutative Yang-Mills theory on D4-brane
from infinitely many D2-branes.
Going back to the original variable $\phi^I$ in (\ref{TrgtX})
for $I = 6,\cdots,10$,
all the terms in (\ref{NBM5H2}) are also finite in the limit \ref{NBM5lim}).

Now, we would like to examine the double dimensional
reduction of the NP M5-brane action.
We compactify the $x^5$ 
 direction with
a coordinate compactification radius  $R_5^{coord}$ 
as in subsection~\ref{secMIIA}.
$R_5^{coord}$ is related to the physical compactification radius $R_5^{phys}$ as
\bea
 \label{R5}
(R_5^{phys})^2 = g_{55} (R_5^{coord})^2 .
\eea
The background $C$-field
is related to the background $B$-field
through the double dimensional reduction:
\bea
 \label{DDRC}
C_{345} (2\pi R_5^{coord}) = B_{34} .
\eea
Then, the non-commutative parameter $\theta^{34}$
and the Nambu-Poisson tensor $\theta^{345}$ are related as
\bea
 \label{DDRth}
\theta^{345} = \theta^{34} (2\pi R_5^{coord}) ,
\eea
by the postulate (\ref{Cth}) with $\theta^{34}$ in (\ref{theta}).
Actually, our postulate (\ref{Cth}) was
made so that it is
consistent with $\theta^{34}$ in (\ref{theta}).
Notice that $(\ref{theta})$
follows from our choice $\Phi = - B$ (\ref{choicePhi})
in the freedom in the description (\ref{OpenFreedom}).
Thus our postulate for the Nambu-Poisson tensor (\ref{Cth}) 
leads to the choice $\Phi = - B$ 
upon the circle compactification of M-theory.
The relation (\ref{DDRth}) also follows
from the double dimensional reduction
of the NP M5-brane action.
This is quite expected
since the multiple M2-brane action
should reduce to 
multiple D2-brane action upon 
circle compactification of M-theory,
and the multiple D2-brane action
naturally leads to 
the non-commutative D4-brane action
with the choice (\ref{choicePhi})
as we have seen 
in the previous subsection \ref{BGID2}.
See appendix \ref{AppDDRth} for the explicit calculations.

Let us examine the dimensional reduction of the 
(inverse) open membrane metric
(\ref{GOM}).
We first examine the metric in (34)-directions.
Using the relations (\ref{R5})-(\ref{DDRth}) as well
as the M-theory -- \IIA relation discussed in subsection~\ref{secMIIA},
we obtain
\bea
G^{ij}_{OM} = G^{ij} \quad (i,j = 3,4),
\eea
where $G^{ij}$ is the inverse open string metric given in (\ref{OSinvmet}).
Notice that 
(\ref{OSinvmet}) also follows from the choice (\ref{choicePhi}).
Thus our postulate 
for the (inverse) open membrane metric (\ref{GOM})
also leads to the choice $\Phi = - B$ (\ref{choicePhi})
upon the circle compactification of M-theory.

Regarding the $G^{55}_{OM}$ component of the (inverse) open membrane metric, 
interestingly we have an
analogue of M-\IIA relation (\ref{MIIA}) for the open string/membrane variables
\cite{Gopakumar:2000ep}:
\bea
\sqrt{\frac{1}{G_{OM}^{55}} (R_5^{coord})^2} = G_s \ell_s.
\eea


\subsection{Relation to the notation in Ref.\cite{Ho:2008ve}}\label{HIMSnotation}

In Ref.\cite{Ho:2008ve},
an explicit parametrization
of the Nambu-Poisson tensor was used. 
While it is convenient for actual calculations,
when using this parametrization
one should keep in mind that 
it is an expression in a particular coordinate system.
In this paper, we keep track of the tensor structures
so that we can
keep manifest covariance, 
in particular, under the rescaling of the coordinates.
This covariant tensor notation  
is useful when
discussing ambiguities in the Seiberg-Witten map,
as we will see in section \ref{secSWmap}.

In the following, we clarify 
the relation of the notation in our paper and that in 
Ref.\cite{Ho:2008ve}.

From (\ref{M5bg}) we have
\bea
\{X_{bg}^3,X_{bg}^4, X_{bg}^5 \} = \theta^{345} .
\eea
In Ref.\cite{Ho:2008ve}, 
the 3-bracket was given as
\bea
 \label{HIMS3bra}
[A,B,C] =
g^2 \ell^3 \e^{ijk} \frac{\pa}{\pa y^i} A(y) \frac{\pa}{\pa y^j} B(y) \frac{\pa}{\pa y^k} C(y), 
\eea
where
\bea
 \label{HIMSx}
x^i = \frac{y^i}{g} .
\eea
In the above, we have introduced a length scale $\ell$
so that the mass dimension of the 3-bracket is zero.
The convention in Ref.\cite{Ho:2008ve} can be regarded
as setting $\ell = 1$.
From (\ref{HIMS3bra}) and (\ref{HIMSx}) we obtain
\bea
[X_{bg}^3,X_{bg}^4, X_{bg}^5 ]
= \frac{\ell^3}{g} .
\eea
This should be compared with the 3-bracket in our convention
(\ref{Our3bra}).
We obtain the relation
\bea
 \label{thHIMS}
\theta^{345} = \frac{\ell^3}{g} .
\eea
Note that 
the expression in (\ref{thHIMS}) is the component of the
Nambu-Poisson tensor in the $x$ coordinates.

The inner product in this paper was given as (\ref{tr3}):
\bea
 \label{tr32}
\int \frac{d^3 x}{2\pi} \frac{1}{|\theta^{345}|} .
\eea
In terms of $y$ coordinates, (\ref{tr32}) becomes
\bea
\int \frac{d^3 y}{2\pi} \frac{1}{g^3 |\theta^{345}|} =
\int \frac{d^3 y}{2\pi} \frac{1}{g^2 \ell^3} .
\eea
This is basically the same with eq.(3.5)
of Ref.\cite{Ho:2008ve}
after setting $\ell =1$, 
up to the convention for the $2\pi$ factor.

In Ref.\cite{Ho:2008ve},
the metric on the M5-brane in (345)-directions was given by 
the Kronecker delta.
This can be achieved by first taking the closed membrane metric as
\bea
 \label{HIMScmet}
g_{ij} = (2\pi)^2\frac{\ell_P^3}{\ell^3} \delta_{ij} \quad (i,j=3,4,5).
\eea
Note that (\ref{HIMScmet}) follows the scaling (\ref{NBM5lim}).
With this choice of closed string metric, 
the (inverse) open membrane metric becomes
\bea
 \label{OMMx}
G_{OM}^{ij} = \frac{1}{g^2} \delta^{ij} .
\eea
Then, via the coordinate transformation from
$x^i$ to $y^i$ as in (\ref{HIMSx}),
the (inverse) open membrane metric becomes the Kronecker delta
in $y$-coordinates:
\bea
 \label{OMMy}
G_{OMy}^{ij} = \delta^{ij},
\eea
where we used the subscript $y$ to indicate that
(\ref{OMMy}) is the component expression in $y$ coordinates.
Since in $y$-coordinates the metric is kept fixed,
it is a convenient coordinate system for
measuring the physical strength of the 
interaction through the NP bracket.

In Ref.\cite{Ho:2008ve}, 
the fields $X^i$ $(i=3,4,5)$ were parametrized as
\bea
 \label{XHIMS}
X^i = \frac{y^i}{g} + \hat{b}^{(g)i}(y)
= \frac{y^i}{g} + \frac{1}{2} \e^{ijk} \hat{b}^{(g)}_{jk}(y), 
\eea
where we put superscript $(g)$ to the corresponding variables
in the notation of Ref.\cite{Ho:2008ve}.
On the other hand, in this paper
they were parametrized as
\bea
 \label{Xc}
X^i = x^i + \hat{b}^i(x) = x^i + \frac{1}{2} \theta^{ijk} \hat{b}_{jk}(x).
\eea
To compare (\ref{Xc}) with (\ref{XHIMS}),
we should first make coordinate transformation
from $x$ to $y$ related by (\ref{HIMSx}):
\bea
 \label{Xcy}
X^i = \frac{1}{g}(y^i + \hat{b}^i(y)) =  
\frac{1}{g} ( y^i + \frac{1}{2} \theta^{ijk}_y \hat{b}_{jk}(y)  ),
\eea
where $\theta^{ijk}_y$ is the component of the
Nambu-Poisson tensor in $y$ coordinates, which from
(\ref{thHIMS}) is given by
\bea
 \label{thetay}
\theta^{345}_y = g^3 \theta^{345} = g^2 \ell^3.
\eea
Note that as contravariant and covariant tensor fields, 
$\hat{b}^i$ and $\hat{b}_{ij}$ change under change of coordinates.
Our notation is that 
we use $\hat{b}^i(x)$ and $\hat{b}^i(y)$ to denote 
the vector field $\hat{b}^i$ in the $x$ and $y$ coordinate systems, 
respectively.
As a result,
\be
\hat{b}^i(y) = g\hat{b}^i(x), \qquad
\hat{b}_{ij}(y) = g^{-2}\hat{b}_{ij}(x).
\ee
Thus we obtain the relation between current convention and
the convention in Ref.\cite{Ho:2008ve}:
\bea
 \label{HIMSbi}
\hat{b}^{(g)i} (y) &=& \frac{1}{g}\hat{b}^i(y) , \nn \\
\hat{b}_{ij}^{(g)}(y) &=& g {\ell^3} \hat{b}_{ij}(y) .
\eea

In the BLG model,
the covariant derivative was given in (\ref{Dmu})
\bea
(D_\mu \varphi)_a = \pa_\mu \varphi_a - f^{bcd}{}_a A_{\mu bc} \varphi_d  .
\eea
This can be rewritten as
\bea
D_\mu \varphi = \pa_\mu \varphi - A_{\mu bc} [ T^b, T^c, \varphi ] .
\eea
When we defined the Lie 3-algebra 
through the Nambu-Poisson structure on ${\mathbb R}^3$,
the elements $T^a$ of the algebra
were given by square-integrable functions 
on ${\mathbb R}^3$.

We define the components of the
two-form field $\hat{b}_{\mu i}$ by
\bea
\hat{b}_{\mu i} \equiv A_{\mu bc} T^b \pa_i T^c 
\quad (\mu = 0,1,2; i = 3,4,5). 
\eea
Then, the covariant derivatives in $(012)$-directions
in the NP M5-brane theory
can be written as
\bea
 \label{Dmub}
D_{\mu} \hat{\varphi}
=
\pa_{\mu}  \hat{\varphi}
- \theta^{ijk} \frac{\pa}{\pa x^i}\hat{b}_{\mu j}(x) \frac{\pa}{\pa x^k} \hat{\varphi} .
\eea
This should be compared with the convention of Ref.\cite{Ho:2008ve}:
\bea
D_{\mu} \hat{\varphi}^{(g)}
=
\pa_\mu
\hat{\varphi}^{(g)}
- g \e^{ijk} \frac{\pa}{\pa y^i} \hat{b}^{(g)}_{\mu j}(y) \frac{\pa}{\pa y^k} \hat{\varphi}^{(g)} .
\eea
Comparing them in the $y$ coordinates, we obtain the relation
\bea
 \label{HIMSbmui}
\hat{b}_{\mu i}^{(g)}(y) = g {\ell^3} \hat{b}_{\mu i}(y).
\eea

\section{An all order solution to the Seiberg-Witten map}\label{secSWmap}

\subsection{Seiberg-Witten map}

Seiberg-Witten map is 
a map between ordinary description and non-commutative description
of a gauge theory, 
determined by the requirement
that the gauge transformation
for the non-commutative description
is induced by the gauge transformation
in the ordinary description.\footnote{%
Essentially the same problem had been considered
in the study of fermions in the lowest Landau level
\cite{Sakita:1993mc}.}
In the original paper by Seiberg and Witten \cite{Seiberg:1999vs},
the map was explained
by identifying the two descriptions as
two different regularizations on the open string worldsheet theory
in a constant $B$-field background.
The difference in the regularization should not
lead to different space-time S-matrices,
and therefore fields in two descriptions should be 
related by field redefinitions.
Later on, the Seiberg-Witten map between
ordinary description and Poisson bracket description was explained as
different gauge fixings of the reparametrization invariance
on the D-brane worldvolume
\cite{Ishibashi:1999vi,Cornalba:1999hn,Cornalba:1999ah}
(see \cite{Okuyama:1999ig,Jurco:2000fb,Jurco:2000fs} 
for a related approach in constructing Seiberg-Witten map 
between ordinary description and Moyal-product description).
We can follow a similar approach for 
constructing the Seiberg-Witten map between
ordinary description and NP description of M5-brane
in a constant $C$-field background.

In the case of M5-brane in a constant $C$-field background,
Seiberg-Witten map is a solution to the condition:
``Gauge transformations in the Nambu description
is compatible with gauge transformations in the ordinary description":
\bea
 \label{SWc}
\hat{\delta}_{\hLambda} \hPhi (\Phi) = 
\hPhi (\Phi + \delta_\Lambda \Phi) - \hPhi (\Phi) ,
\eea
where $\hPhi$ ($\Phi$) collectively represents
fields in the NP bracket (ordinary) description of M5-brane.
The gauge transformation laws
in the NP M5-brane theory were derived in \cite{Ho:2008ve},
which together with those in the ordinary description of M5-brane
we summarize in our notation below.
The gauge transformation laws
of $b_{ij}$ and $\hat{b}_{ij}$ are given by
\bea
 \label{gtrbij}
\delta_{\Lambda} b_{ij}
&=&
\pa_i {\Lambda}_j - \pa_j {\Lambda}_i , \nn \\ 
\delta_{\hat{\Lambda}} \hat{b}_{ij} 
&=&
\pa_i \hat{\Lambda}_j - \pa_j \hat{\Lambda}_i 
+
\theta^{\ell mn} \pa_\ell \hat{\Lambda}_m \pa_n \hat{b}_{ij} . 
\eea
In terms of 
${b}^i \equiv \frac{1}{2} \theta^{ijk} {b}_{jk}$ and
$\hat{b}^i \equiv \frac{1}{2} \theta^{ijk} \hat{b}_{jk}$, these can be rewritten as
\bea
 \label{gtrbi}
\delta_{{\Lambda}}  {b}^i = {\kappa}^i  , \qquad
\delta_{\hat{\Lambda}}  \hat{b}^i = \hat{\kappa}^i + \hat{\kappa}^j \pa_j \hat{b}^i ,
\eea
where
\bea
 \label{kappas}
{\kappa}^i \equiv \theta^{ijk} \pa_j {\Lambda}_k , \qquad
\hat{\kappa}^i \equiv \theta^{ijk} \pa_j \hat{\Lambda}_k .
\eea
From (\ref{kappas}), the gauge transformation parameters
$\kappa^i$ and $\hat{\kappa}^i$ satisfy the divergenceless condition
\bea
\pa_i \kappa^i = 0, \qquad \pa_i \hat{\kappa}^i =0.
\eea
Note that $\hat{b}^i$ and $\hat{\kappa}^i$ are 
one order higher in the expansion in $\theta$
compared with $\hat{b}_{ij}$ and $\hat{\Lambda}_i$, respectively.
The gauge transformation laws of 
${b}_{\mu i}$ and $\hat{b}_{\mu i}$ are given by
\bea
 \label{gtrbmi}
\delta_{\Lambda} {b}_{\mu i} 
&=&
\pa_{\mu} \Lambda_{i} - \pa_i \Lambda_{\mu}  ,
\nn \\
\delta_{\hat{\Lambda}} \hat{b}_{\mu i} 
&=&
\pa_{\mu} \hat{\Lambda}_{i} - \pa_i \hat{\Lambda}_{\mu}
+ \hat{\kappa}^{j}\pa_j \hat{b}_{\mu i}
+ \pa_i \hat{\kappa}^j \hat{b}_{\mu j} .
\eea
The gauge transformation laws for $\varphi$ and $\hat{\varphi}$
are given by
\bea
 \label{gtrvphi}
\delta_{{\Lambda}} {\varphi} = 0, \quad 
\delta_{\hat{\Lambda}} \hat{\varphi}
&=&\hat{\kappa}^j \pa_j \hat{\varphi} . 
\eea

The Seiberg-Witten map to the first order was obtained in \cite{Ho:2008ve} as
\bea
\hat{b}^i &=& b^i + \frac{1}{2} b^j\pa_j b^i + \frac{1}{2} b^i \pa_j b^j 
+ {\cal O}(\theta^3),\\
\hat{B}_{\mu}{}^i &=& B_{\mu}{}^i + b^j \pa_j B_{\mu}{}^i 
- \frac{1}{2} b^j \pa_{\mu} \pa_j b^i + \frac{1}{2} b^i \pa_{\mu} \pa_j b^j
+ \pa_j b^j B_{\mu}{}^i - \pa_j b^i B_{\mu}{}^j \nn \\
&& - \frac{1}{2} \pa_j b^j \pa_{\mu} b^i + \frac{1}{2} \pa_j b^i \pa_{\mu} b^j 
+ {\cal O}(\theta^3),\\
\hat{\kappa}^i &=& \kappa^i + \frac{1}{2} b^j \pa_j \kappa^i + \frac{1}{2} \pa_j b^j \kappa^i
- \frac{1}{2} \pa_j b^i \kappa^j 
+ {\cal O}(\theta^3), \\
\hat{\varphi} &=&
\varphi + \kappa^{i} \pa_i \varphi + \Ord (\theta^2) .
\eea
Here we have used $\hat{B}_{\mu}^i$ (and $B_{\mu}^i$) 
defined below
in place of $\hat{b}_{\mu i}$ (and $b_{\mu i}$)
to express the Seiberg-Witten map.
As mentioned earlier, in the expansion in $\theta$
we regard ${b}_{ij}, {b}_{\mu i}$, $\varphi$ and ${\Lambda}_i$ as $\Ord (1)$
variables,
and thus count ${b}^i, {B}_{\mu}{}^i$ and ${\kappa}^i$ as $\Ord (\theta)$.
Thus, above expansion corresponds to the first order in $\theta$ expansion
in terms of ${b}_{ij}, {b}_{\mu i}$, $\varphi$ and ${\Lambda}_i$.

In the covariant derivatives and the action, 
$\hat{b}_{\mu i}$ always appears in the form 
\be
 \label{BhB}
\hat{B}_{\mu}{}^i \equiv \theta^{ijk}\pa_j \hat{b}_{\mu k},
\ee
and thus we can avoid the explicit use of $\hat{b}_{\mu i}$ 
by using $\hat{B}_{\mu}{}^i$ 
with the constraint
\be
\pa_i \hat{B}_{\mu}{}^i = 0, 
\ee
which guarantees the existence of $\hat{b}_{\mu i}$ locally, 
and similarly for the variables without hats.
For various purposes, $\hat{B}_{\mu}{}^i$ is more convenient to use.
The gauge transformation laws of $\hat{B}_{\mu}{}^i$ and $B_{\mu}{}^i$ are given by
\bea
\delta_{\Lambda} B_{\mu}{}^i &=& \pa_{\mu}\kappa^i, \\
\hat{\delta}_{\hat{\Lambda}} \hat{B}_{\mu}{}^i &=& \pa_{\mu}\hat{\kappa}^i
+ \hat{\kappa}^j \pa_j \hat{B}_{\mu}{}^i - \pa_j \hat{\kappa}^i \hat{B}_{\mu}{}^j.
\eea
It turns out that it is easier to write down the Seiberg-Witten map for
$\hat{B}_{\mu}{}^i$ than the one directly relating $\hat{b}_{\mu i}$ to $b_{\mu i}$.

To obtain an all order solution
to the Seiberg-Witten map,
we follow the approach of \cite{Jurco:2000fb,Jurco:2000fs}
which was applied to the
Poisson bracket $U(1)$ gauge theory.
We will generalize their construction
to the NP M5-brane theory.
However in the case of NP M5-brane theory,
there is the new ingredient that
(012)-directions and (345)-directions are 
related through the self-dual relations 
for the two-form gauge field.

%

\subsection{Nambu-Poisson manifold}

To construct an all order solution to the Seiberg-Witten map,
we need to extend our consideration to 
include general Nambu-Poisson manifold \cite{Takhtajan}.
A Nambu-Poisson manifold
is defined through a Nambu-Poisson bracket
which is tri-linear and totally skew-symmetric in its entries:
\bea
 \label{oNB}
\{ A,B,C \} = 
\theta^{ijk}(x) \frac{\pa}{\pa x^i} A(x) \frac{\pa}{\pa x^j} B(x) \frac{\pa}{\pa x^k} C(x),
\quad (i,j = 1,2,3),
\eea
and which satisfies the {\em fundamental identity}:
\bea
 \label{FI2}
\{A,B,\{C,D,E \} \}
=
\{ \{A,B,C \} D, E \} + \{ C, \{A,B,D\} ,E \} + \{ C,D, \{A,B,E \} \} .
\eea
The fundamental identity (\ref{FI2}) puts strong constraints on 
the totally anti-symmetric Nambu-Poisson tensor $\theta^{ijk}(x)$.
For example, the following identities hold (see e.g. \cite{survey}):
\bea
\theta^{ib_2b_3} \theta^{ja_2b_1} 
+
\theta^{b_1 i b_3} \theta^{ja_2b_2} 
+
\theta^{b_1 b_2 i} \theta^{j a_2 b_3} 
+ (i \leftrightarrow j)
= 0,
\eea
\bea
\theta^{a_1a_2 i} \pa_i \theta^{b_1b_2b_3}
=
\theta^{ib_2b_3}\pa_i \theta^{a_1a_2b_1}
+
\theta^{b_1ib_3} \pa_i \theta^{a_1a_2b_2}
+
\theta^{b_1b_2i} \pa_i \theta^{a_1a_2b_3}
.
\eea
For a given 
Nambu-Poisson tensor ${\theta'}^{ijk}$,
what we would like to have is 
a coordinate transformation $\rho(x^i)$:
\bea
\rho^* {\theta'}^{ijk} = \theta^{ijk} ,
\eea
i.e. a coordinate transformation
that maps back the Nambu-Poisson bracket 
defined by ${\theta'}^{ijk}$
to the original Nambu-Poisson bracket (\ref{oNB}) defined by $\theta^{ijk}$,
\bea
\rho^* \{A,B,C \}' = \{\rho^* A, \rho^* B , \rho^* C \} ,
\eea
where $\{ * ,* , *\}' $ is the Nambu-Poisson bracket
defined by the Nambu-Poisson tensor ${\theta'}^{ijk}$.
One can construct such a 
map by using a flow parametrized by $t$:
\bea
 \label{floweq}
\pa_t \{A,B,C \} + \chi(t) \{A,B,C\}
- \{ \chi (t) A ,B,C \} - \{ A, \chi (t) B, C \} - \{ A, B, \chi (t) C \} = 0,
\eea
where 
\bea
 \label{chi}
\chi \equiv \frac{1}{2}\theta^{ijk}(t) b_{ij} \pa_k ,
\eea 
and
\bea
 \label{inith}
\theta^{ijk} (t=0) = \theta^{ijk}, \quad \theta^{ijk}(t=1) = {\theta'}^{ijk} .
\eea
The tensor $b_{ij}$ in (\ref{chi}) is related to ${\theta'}^{ijk}$ as
\bea
-(C+H)_{ijk} {\theta'}^{k\ell m} = \delta_i^\ell \delta_j^m -  \delta_i^m \delta_j^\ell ,
\eea
where
\bea
H_{ijk} \equiv \pa_i b_{jk} + \pa_j b_{ki} + \pa_k b_{ij} .
\eea
Using the fundamental identity (\ref{FI2}),
the flow equation (\ref{floweq}) can be written
as a first order differential equation for $\theta^{ijk}(t)$:
\bea
 \label{floweqt}
\pa_t \theta^{ijk} (t)
&=&
\frac{1}{6}
\left(
\theta^{a_1 ij} (t) \theta^{a_2a_3 k} (t)
+
\theta^{a_1 jk} (t) \theta^{a_2a_3 i} (t)
+
\theta^{a_1 ki} (t) \theta^{a_2a_3 j} (t)
\right)
H_{a_1 a_2 a_3} \nn\\
&=&
\frac{1}{6}
\theta^{ijk}(t) \theta^{a_1a_2a_3}(t) H_{a_1a_2a_3}.
\eea
with the initial condition at $t=0$ as in (\ref{inith}).
The explicit solution for (\ref{floweqt}) 
with the initial condition mentioned above is given by
\bea
\theta^{ijk}(t)
= 
\theta^{ijk} \frac{1}{1 - \frac{t}{6} (\theta^{a_1a_2a_3} H_{a_1a_2a_3})} .
\eea
%

\subsection{An all order solution to the Seiberg-Witten map}

An all order solution 
to the Seiberg-Witten map 
can be constructed using
the flow discussed 
in the previous subsection.
We first construct the Seiberg-Witten map
for the fields $b^i$ and $\hat{b}^{i}$.
%
Our
solution to the Seiberg-Witten map is 
as follows:
\bea
  \label{allordx}
\rho(x^i) = 
x^i + \hat{b}^i =
e^{\pa_t + \frac{1}{2} \theta^{ijk}(t) b_{ij} \pa_k }  \,
x^i  
\Bigr|_{t=0}  .
\eea
We first check that 
(\ref{allordx}) leads to the correct 
infinitesimal gauge transformations (\ref{gtrbi}).
It is also possible to write down the Seiberg-Witten map 
for finite gauge transformations,
as opposed to infinitesimal gauge transformations 
considered here. 
We derive 
the Seiberg-Witten map 
for 
finite gauge transformation parameters in the appendix \ref{finiteGT}.

From the map (\ref{allordx}), the infinitesimal gauge transformation
in the NP description 
induced by the infinitesimal gauge transformation
of the ordinary description
is given by
\bea
&& \hat{b}^{i}(b + \delta_{\Lambda} b)
- \hat{b}^{i}(b) \nn \\
&=& 
\left[
e^{\pa_t + \frac{1}{2}\theta^{ijk}(t) (b_{ij}+\pa_i \Lambda_j - \pa_j \Lambda_i) \pa_k }
-
e^{\pa_t + \frac{1}{2}\theta^{ijk}(t) b_{ij} \pa_k }  \right]
x^i  
\Bigr|_{t=0}  .
\label{allorderg}
\eea
Below we demonstrate that (\ref{allorderg})
indeed gives a solution to the Seiberg-Witten map,
i.e. satisfies the condition (\ref{SWc}).
Recall that the gauge transformation laws 
for the fields $b^i$ and $\hat{b}^{i}$
are given in (\ref{gtrbi}).
Let us write
\bea
 \label{AB}
A \equiv \pa_t + \frac{1}{2} \theta^{ijk}(t) b_{ij} \pa_k ,
\quad
B \equiv \frac{1}{2} \theta^{ijk}(t) (\pa_i \Lambda_j - \pa_j \Lambda_i) \pa_k ,
\eea
Using this notation, (\ref{allorderg}) 
can be rewritten as
\bea
 \label{eABx}
&&\left( e^{A+B} x^i - e^A x^i  \right) \Bigr|_{t=0} 
=
\left(
\left[
e^{A+B} e^{-A} - 1
\right] 
e^A x^i \right)
\Bigr|_{t=0} .
\eea
Eq. (\ref{eABx})
involves the quantity 
\bea
 \label{eAB}
e^{A+B} e^{-A} 
= e^{h(A,B)}
,
\eea
where the $h(A,B)$ is the linear combination of the terms derived by the 
Baker-Campbell-Hausdorf formula.
In the case of infinitesimal gauge transformation, 
i.e. infinitesimal $\Lambda$, 
among the terms in $h(A,B)$ 
only 
the terms linear in $B$ is relevant.
Such term
has the following form:
\begin{equation}
[\cdots [A, [A, B]] \cdots] .
\end{equation}
Noting that
the $\pa_t$ term in $A$ has a constant coefficient
and there is no $\pa_t$ term in $B$,
(\ref{eAB}) can be written as 
\bea
\label{eABk}
e^{A+B} e^{-A} - 1
=  \hat{\kappa}^i(t) \pa_i + \Ord (\Lambda^2). 
\eea
Using (\ref{eABk}), (\ref{allorderg}) can be written as
\bea
\hat{b}^i(b+ \delta_{\Lambda}b ) - \hat{b}^i(b) = \hat{\kappa}^i + \hat{\kappa}^j \pa_j \hat{b}^i ,
\eea
where we identify the gauge transformation parameter
in the NP description as
\bea
\hat{\kappa}^i \equiv \hat{\kappa}^i(t=0) .
\eea
In the appendix \ref{dlessk}, we show that the gauge transformation parameter
$\hat{\kappa}^i$
defined in (\ref{allorderg}) satisfies the divergenceless condition
\bea
\pa_i \hat{\kappa}^i = 0.
\eea
Therefore, it can be written as $\hat{\kappa}^i = \theta^{ijk} \pa_j \hat{\Lambda}_k$.
Thus, $\hat{b}^i$ given by (\ref{allordx})
is a solution of the Seiberg-Witten map (\ref{SWc}).


Next, let us construct the Seiberg-Witten
map for fields $\varphi$ and $\hat{\varphi}$.
The gauge transformation laws for
$\varphi$ and $\hat{\varphi}$ are given in (\ref{gtrvphi}). 
A solution to the Seiberg-Witten map can be obtained as
\bea
 \label{allordphi}
\hat{\varphi} =
e^A \varphi 
\Bigr|_{t=0}.
\eea
By a calculation similar to the case in $\hat{b}^i$,
one can show that $\hat{\varphi}$ defined in (\ref{allordphi})
is a solution to the Seiberg-Witten map (\ref{SWc}).

The gauge transformation law for $\hat{b}_{\mu i}$ 
is given in (\ref{gtrbmi}). 
To obtain a solution to the Seiberg-Witten map,
it is useful to notice 
that $\hat{b}_{\mu i}$ appears in the covariant derivative (\ref{Dmub}).
Therefore, we first look for a differential operator
in the commutative description
which (i) when acted on a scalar, the covariant derivative of
the scalar transforms as a scalar under the volume-preserving diffeomorphisms;
(ii) contains $b_{\mu i}$ linearly.
Then, we consider an exponential map
similar to the Seiberg-Witten map for the scalar fields.
In this way, we make the following 
guess 
for the solution to the
Seiberg-Witten map:
\bea 
 \label{allordB}
&&
\hpa_{\mu}
-
\hat{B}_\mu{}^i
\hpa_i \nn \\
&=&
e^{A}  
\left(
\hpa_\mu - \theta^{ijk}(t) 
 \left(
\hpa_j b_{\mu k} -\frac{1}{2}\pa_\mu b_{jk}
 \right)\hpa_i
\right)
e^{-{A}}  
\Bigr|_{t=0} .
\eea
where we use $\hpa$ to denote differential {\em operators}
emphasizing
that the differential acts on the whole objects 
in the right, 
for example
\bea
\hpa_t t = t \hpa_t + 1, \quad \hpa_i x^j = x^j \hpa_i + \delta_i^j , \quad
\hpa_i f(x) = \pa_i f(x) + f(x) \hpa_i.
\eea
(\ref{allordB}) turns out to be
indeed a solution to the Seiberg-Witten map.
In appendix \ref{dlessB},
we show that $\hat{B}_\mu{}^i$ satisfies the divergenceless condition:
\bea
 \label{divlessB}
\pa_i \hat{B}_\mu{}^i =0,
\eea
and hence can be written as
\bea
 \label{hBhb}
\hat{B}_\mu{}^i = \theta^{ijk} \pa_j \hat{b}_{\mu k} .
\eea
On the other hand,
we can show the gauge transformation law
for $\hat{B}_\mu{}^i$ in a way similar
to the previous cases.

Our solution of the Seiberg-Witten map
(\ref{allordx}), (\ref{allordphi}) and (\ref{allordB})
are expressed in a covariant tensor notation,
and hence does not depend on particular coordinates.
On the other hand,
when one would like to study the small Nambu-Poisson tensor expansion,
one should fix the open membrane metric as in (\ref{OMMy})
to compare the physical strength of the interaction
through the Nambu-Poisson structure.
Then we can more precisely
specify the expansion as one in the
small Nambu-Poisson tensor component in the $y$-coordinates
(\ref{thetay}),
when we take the closed membrane metric as in (\ref{HIMScmet}).

\subsection{Ambiguities in the SW map}

The Seiberg-Witten map is not unique \cite{Asakawa:1999cu,Okawa:2000em}. 
There are two sources of ambiguities.
The first one arises from the fact
that the Seiberg-Witten map is
basically a map between gauge orbits in two descriptions of the same theory.
Therefore, replacing $b_{ij}$ in the expression (\ref{allordx}) by
$b_{ij} + \delta_{\Lambda} b_{ij}$ 
for any gauge transformation $\Lambda$, for example, 
gives another Seiberg-Witten map, 
because a gauge transformation on $b_{ij}$ 
does not affect the identification of the gauge orbit.
%
%
Even if we insist that the Seiberg-Witten map should not involve 
anything other than $b^i$, $\theta^{ijk}$, $C_{ijk}$ and $\pa_i$,
we can still get a new Seiberg-Witten map by taking 
$\Lambda_i = C_{ijk} b^j \pa_l b^k b^l$, for instance.
This is the only ambiguity at the order of ${\cal O}(\theta^2)$
under such assumptions,
and there is no ambiguity of this kind at lower orders.
But apparently at higher orders there are more and more such ambiguities.
In general, the SW map for $\hat{b}^i$, for example,
can be of the form
\be
\hat{b}^i =
\left(e^{\pa_t + \frac{1}{2} \theta^{ijk}(t)(b_{ij}+\pa_i \Lambda_j - \pa_j \Lambda_i) \pa_k } - 1\right)  \,
x^i \Bigr|_{t=0} 
\ee
for some vectors $\Lambda_i(b, \pa b, \cdots)$
which are given functions of the dynamical fields.

The second source of ambiguities comes from 
field redefinitions.
A simple way to see the potential existence of such ambiguities is the following.
If we try to solve for the SW map order by order from 
the defining condition (\ref{SWc}),
we can always add a gauge invariant term 
with appropriate 
Lorentz transformation property 
at the $n$-th order in $\theta^{ijk}$ 
without spoiling the condition (\ref{SWc}) at lower orders.
Then we can solve for higher order terms of the SW map 
accordingly.
The all order solutions of this type are of the form
\be
\hat{b}^i =
\left(e^{\pa_t + \frac{1}{2} \theta^{ijk}(t) b_{ij} \pa_k } - 1\right)  \,
x^i \Bigr|_{t=0}
+ e^{\pa_t + \frac{1}{2} \theta^{ijk}(t) b_{ij} \pa_k } f^{i}(H,\pa H, \pa^2 H, \cdots)
\Bigr|_{t=0},
\ee
where $f^i$ is a 
local gauge invariant vector.
As $f^i$ is 
gauge 
invariant, such modifications of the SW map 
does not change the SW map for the gauge transformation parameter
$\hat{\kappa}^i$ (\ref{eABk}). 

As an example, one can choose
\be
f^i = \sum_{r=0}^{\infty} c_r G^r \theta^{ijk}\theta^{lmn} \pa_l H \pa_j\pa_m H \pa_k\pa_n H,
\ee
where $c_r$ are arbitrary numbers and 
\be
H \equiv \theta^{ijk} H_{ijk}, \qquad
G \equiv \theta^{ijk}\theta^{lmn} \pa_i\pa_l H \pa_j\pa_m H \pa_k\pa_n H.
\ee
This example is constructed such that the coefficients $c_r$ are all dimensionless 
and we have only used the minimal set of variables, i.e., 
the invariant tensor $H_{ijk}$ and $\theta^{ijk}$.
Furthermore $f^i$ is not divergenceless, 
so even at the lowest nontrivial order it can not be mistaken as 
the other kind of ambiguity due to a gauge transformation.

Since field redefinitions change the form of the action,
this type of ambiguities may be fixed
by restricting the form of the action \cite{Okawa:2000em}.
Notice that 
our 
covariant tensor notation 
is useful for writing down all possible terms
in the discussions of ambiguities.

Our solution to the Seiberg-Witten map
has the clear meaning 
that it is a map between
two different choices of coordinates,
one which keeps the NP structure and 
the other so-called static gauge.
It will be interesting to investigate further
to what extent our solution is special
among all possible solutions of Seiberg-Witten map. 


\section{Discussions}\label{secDiscussions}

In this paper, 
we obtained several results
which will be essential for showing
the conjectured equivalence of the
conventional M5-brane theory in a constant $C$-field background
and NP M5-brane theory.
The scaling limit discussed in section \ref{secNBM5lim}
is necessary for identifying the 
region of validity of the NP M5-brane theory,
and it also specifies how to take a limit
in the conventional M5-brane theory
in order to compare the two theories.
The precise identification of the variables in the NP M5-brane theory
with 
M-theory variables 
is another necessary step to relate the two M5-brane theories.
%
The background-independent formulation
was useful for reading off the open membrane metric and effective tension of the
NP M5-brane.
The all-order solution to the Seiberg-Witten map
is of course essential for establishing 
the relation between two M5-brane theories.
The logical steps we took to find the solution
already indicates the
equivalence of the two descriptions,
as different choices of variables of the same theory.
In the meantime,
as pointed out in Ref.\cite{Ho:2008ve}
the field $\hat{B}_\mu{}^i$
has similarity with 
the gauge field parameterizing complex structure
deformations on a Calabi-Yau 3-manifold
in the Kodaira-Spencer theory,
and we feel our derivation has a room for
mathematical sophistications.

Motivated by \cite{Ho:2008ve},
a new covariant action for
the self-dual 2-form gauge field 
which is based on the 3+3 decomposition
of the worldvolume was formulated
at the linear level in \cite{Pasti:2009xc}
(see \cite{Chen:2010jg} for a generalization to
chiral p-form gauge field with general decomposition of the space-time).
Since NP M5-brane theory is based on the 3+3 decomposition
of the worldvolume,
it seems more convenient to 
reformulate the conventional M5-brane theory
also in the 3+3 decomposition of the worldvolume.
Investigation in this direction will also be useful.

\section*{Acknowledgments}

The authors thank 
Wei-Ming Chen,
Andreas Gustavsson,
Takeo Inami,
Hiroshi Isono,
Sheng-Lan Ko,
Yutaka Matsuo,
Shin Sasaki,
Peter Schupp,
Shotaro Shiba,
Sheng-Yu Darren Shih,
Seiji Terashima
and Dan Tomino
for helpful discussions.
KF thanks KEK for the invitation to 
the KEK theory workshop 2010 
where some results
in this paper were announced.
This work is supported in part 
by the National Science Council,
and the National Center for Theoretical Sciences, Taiwan, R.O.C.

\appendix
\section{Absence of $\theta^{012}$
in the NP M5-theory scaling limit}\label{AppScaling}
We consider the scaling limit to the
NP M5-brane theory (\ref{NBM5lim}):
\bea
\label{AppNBM5lim}
\quad \ell_P &\sim& \e^{1/3}, \nn \\
g_{ij} &\sim& \e, \quad (i,j = 3,4,5) \nn \\
\quad C_{345} &\sim& \e^0 .
\eea
We define
the (inverse) open membrane metric
in $(012)$-directions
in a similar way as in (\ref{GOM}):
\bea
 \label{OMM0}
G^{\mu\mu'}_{OM}
\equiv 
\frac{1}{2}
\frac{\theta^{\mu\nu\rho}}{(2\pi)^2\ell_P^3}
\frac{\theta^{\mu'\nu'\rho'}}{(2\pi)^2\ell_P^3}
g_{\nu\nu'}g_{\rho\rho'}
\quad
(\mu,\nu,\cdots = 0,1,2).
\eea
Here, as in (\ref{Cth}), we postulate
that the Nambu-Poisson tensor $\theta^{\mu\nu\rho}$
is given by
\bea
 \label{Ct0}
-C_{\mu\nu\rho} \theta^{\rho\b\gamma} =
\delta_{\mu}^{\b}\delta_{\nu}^{\gamma} 
- \delta_{\mu}^{\gamma}\delta_{\nu}^{\b} .
\eea
As discussed in section \ref{secNBM5lim}, 
from the non-linear self-dual relation
we obtain the scaling
\bea
 \label{C012}
C_{012} \sim \e^{-1} ,
\eea
with the closed membrane metric scaling as
\bea
\label{cm0}
g_{\mu\nu} \sim \e^0 .
\eea
From (\ref{Ct0}) and (\ref{C012}),
we obtain 
\bea
\theta^{012} \sim \e ,
\eea
and thus $\theta^{012}$ vanishes in the scaling limit $\e \rightarrow 0$. 
Notice that this is the result
in which the (inverse) open membrane metric
(\ref{OMM0}) is finite,
thus it correctly measures the strength
of the interaction through $\theta^{012}$.

Let us study the compactification in $x^2$ direction 
to see, via the M-\IIA relation, 
what the above scaling corresponds to
in type \IIA string theory.
We take the scaling of the 
physical compactification radius $R_2^{phys}$ as
\bea
\label{R2p}
R_2^{phys} \sim \e^a .
\eea
Then, from (\ref{cm0}),
the coordinate compactification radius $R_2^{coord}$ also scales as
\bea
\label{R2c}
R_2^{coord} \sim \e^a ,
\eea
where the number $a$ is to be determined.
The scalings of the type \II A variables
are given as
\bea
\label{IIAv0}
\ell_s &=&
\left(
\frac{\ell_P^3}{R_2^{phys}} 
\right)^{1/2}
\sim \e^{\frac{1-a}{2}},
\nn \\
g_s &=&
\left(
\frac{R_2^{phys}}{\ell_P}
\right)^{3/2}
\sim 
\e^{\frac{3a-1}{2}} .
\eea
From
(\ref{IIAv0}) we should assume $a < 1$ to take the zero-slope limit
in type \IIA string theory.
The $B$-field is related to the $C$-field as
\bea
B_{01} = C_{012} (R_2^{coord}) \sim \e^{-1+a} .
\eea
Then from (\ref{IIAv0})
\bea
2\pi\a' B_{01} \sim \e^0.
\eea
We will not take the OM-theory limit \cite{Gopakumar:2000ep}
(which is given by ${(1-2\pi\a' B_{01})}/{\a'} \sim \e^0$).
Then, the open string metric in (\ref{OSmet}) and the 
non-commutative parameter $\theta^{01}$
in (\ref{theta})
scale as 
\bea
G_{\mu\nu} &\sim& \e^0, \\
\theta^{\mu\nu} &\sim& \a' \sim \e^{1-a}, 
\quad (\mu,\nu = 0,1).
\eea
Thus in the scaling limit $\e \rightarrow 0$,
there is no interaction through 
the Poisson bracket in $(01)$-directions
as long as $a < 1$.

As noted in subsection \ref{BGINP5},
our postulates
for the open membrane metric
and the Nambu-Poisson tensor
reduces to the 
$\Phi=-B$ description (\ref{choicePhi})
upon circle compactification.
On the other hand,
when
$2\pi\a' B_{\mu\nu} \ll g_{\mu\nu}$,
it may be better to use 
$\Phi =0$ description 
in (\ref{OpenFreedom})
instead of $\Phi = -B$
description above 
so that one can treat the 
open membrane metric as a
small deformation from the closed string metric.
The conclusion that
$\theta^{01} \sim \e^{1-a}$ does not change even in this case.
On the other hand,
we currently do not know
how to uplift the freedom in the description
in (\ref{OpenFreedom}) to M-theory.

We may formally require
the Yang-Mills coupling (\ref{scl})
on the D4-brane 
to be finite,
although it is actually $U(1)$ gauge theory:
\bea
g_s \sim \e^{\frac{3-p+r}{4}} .
\eea
Here, $p=4$ and $r$ is the rank of the Poisson tensor 
finite in the scaling limit, which is zero as above.
Comparing with (\ref{IIAv0}),
we have
\bea
a = \frac{1}{6}.
\eea

On the other hand,
in this compactification
the interaction through the 
Nambu-Poisson bracket in $(345)$-directions
is interpreted as induced on D4-brane by
RR 3-form flux in type \IIA string theory.
Such a system has not been studied much previously.
If we require finite interaction through
the Nambu-Poisson bracket in $(345)$-directions
in this theory, we obtain 
\bea
a=0.
\eea

\section{(\ref{DDRth}) from the double dimensional reduction
of NP M5-brane action}\label{AppDDRth}

It has been shown in \cite{Ho:2008ve}
that by the double dimensional reduction
the NP M5-brane action reduces to the
Poisson description
of a D4-brane in a constant $B$-field background.
Here, it will be enough
to study the 
potential term of the
NP M5-brane theory
including $X^5$ given by
(see (\ref{potX})
and (\ref{tr3})):
\bea
 \label{Apppot}
&&
\frac{1}{(2\pi)^{2}\ell_P^3}
\int d^3 x 
\int \frac{d^3 x}{2\pi |\theta^{345} |} \,
\frac{1}{(2\pi)^{4}\ell_P^6}
\frac{1}{12}
g_{II'}g_{JJ'}g_{KK'}\,
\{ X^I,X^J,X^K \} , \{ X^{I'},X^{J'},X^{K'} \} 
, \nn \\
&& \qquad (I,J = 5,\cdots,10) .
\eea
The double dimensional reduction 
is described by taking
\bea
 \label{DDR}
X^5 = x^5,
\eea
where the coordinate compactification radius 
in the $x^5$ direction is $R_5^{coord}$.
All the other fields are set to be independent of $x^5$.
Then, the potential term (\ref{Apppot}) becomes
\bea
 \label{AppDDRpot}
&&
\frac{1}{(2\pi)^{2}\ell_P^3}
\int d^3 x 
\int \frac{d^2 x \, 
(2\pi R_5^{coord})}{2\pi |\theta^{345} |} \,
\frac{1}{(2\pi)^{4}\ell_P^6}
\frac{1}{4}
g_{II'}g_{JJ'}g_{55}\,
(\theta^{ij5} \pa_i X^I \pa_j X^J  )
(\theta^{i'j'5} \pa_{i'} X^{I'} \pa_{j' }X^{J'} )
, \nn \\
&& \qquad (I,J = 6,\cdots,10) .
\eea
On the other hand,
the potential term of the Poisson description
of D4-brane is given by
(see (\ref{D2pot}) and (\ref{trace}))
\bea
\label{AppD4pot}
&&\frac{1}{(2\pi)^2 g_s \ell_s^3}
\int d^3 x 
\int \frac{d^2 x}{2\pi |\theta^{34}|}\,
\frac{1}{(2\pi\a')^2}
\frac{1}{4}
g_{II'}g_{JJ'}\,
\{ X^I,X^J \}_{Poisson} \{ X^{I'},X^{J'} \}_{Poisson} ,\nn\\
&& \qquad  (I,J = 6,\cdots,10) .
\eea
where $\{ *,* \}_{Poisson}$ is the Poisson bracket (\ref{PB}):
\bea
\{ A,B \}_{Poisson} =
\theta^{ij}
\pa_i A \pa_j B ,\quad (i,j = 3,4) .
\eea
Comparing (\ref{AppDDRpot}) and (\ref{AppD4pot})
using the M-\IIA relation (\ref{MIIA}),
we obtain
\bea
\theta^{34} = \frac{\theta^{345}}{2\pi R_5^{coord}},
\eea
which coincides with (\ref{DDRth}).

\section{Divergenceless condition for $\hat{\kappa}^i$}\label{dlessk}
To make calculation simpler, we use 
the following explicit parametrization of $\theta^{ijk}$:
\bea
 \label{thTh}
\theta^{ijk} = \Theta \e^{ijk} ,
\eea
where $\e^{ijk}$ is a totally anti-symmetric tensor with $\e^{345} = 1$.
Similarly, we parametrize $\theta^{ijk}(t)$ in (\ref{floweqt}) as
\bea
 \label{thTht}
\theta^{ijk}(t) = \Theta(t) \e^{ijk} .
\eea
In this parametrization, (\ref{floweqt}) takes the form
\bea
 \label{gteq}
\pa_t \Theta(t) = \Theta^2(t) \pa\cdot b
\eea
with $\Theta(0) = \Theta$.


$A$ and $B$ in (\ref{AB}) are now written as 
\bea
 \label{gAB}
A = {\pa_t} + \Theta(t) b^i {\pa}_i, \quad
B = \Theta(t) \kappa^i \pa_i,
\eea
where
\bea
\label{appdefbk}
b^i \equiv \frac{1}{2}\e^{ijk}b_{jk}, \quad
\kappa^i \equiv \e^{ijk} \pa_j \Lambda_k .
\eea
In this appendix 
we use notation 
for $b$ and $\kappa$ different from 
the main body (\ref{gtrbi}) by the overall scaling.
From the definition (\ref{appdefbk}),
$\kappa^i$ satisfies the divergenceless condition
\bea
\pa_i \kappa^i = 0.
\eea

Let us first calculate $[A,B]$:
\bea
&&[A,B] \nonumber \\
&=&
\left[
\pa_t + \Theta(t) b \cdot {\pa} \, , \,
\Theta(t) \kappa \cdot \partial
\right] \nonumber \\
&=&
\partial_t \Theta(t) \kappa \cdot {\pa}
+
\Theta(t)(b\cdot \partial \Theta(t) ) \kappa \cdot \pa
+
\Theta^2(t) b\cdot \partial \kappa \cdot \pa
-
\Theta(t) \kappa \cdot \partial (\Theta(t) b) \cdot {\pa} 
\nonumber \\
&=&
\partial_t \Theta(t) \kappa \cdot {\pa}
+
\Theta(t)
\left( 
\partial \cdot (b  \Theta(t) ) 
- (\partial \cdot  b ) \Theta(t)
\right)
\kappa \cdot {\pa} \nonumber \\
&&
+
\Theta(t)^2 b\cdot \partial \kappa \cdot {\pa}
-
\Theta(t) \kappa \cdot \partial (\Theta(t) b) \cdot {\pa} 
\label{usegteq}\\
&=&
\Theta(t) \partial \cdot ( b \Theta(t) )  \kappa \cdot {\pa}
+
\Theta^2(t) b\cdot \partial \kappa \cdot {\pa}
-
\Theta(t) \kappa \cdot \partial (\Theta(t) b) \cdot {\pa}
\nonumber \\
&=&
\Theta(t) \partial_{i}
\left(
\Theta(t) b^{i} \kappa^{j}
-
\Theta(t) b^{j} \kappa^{i}
\right)
{\pa}_{j}
\equiv 
\Theta(t) \tilde{\kappa}_{(1)}(t) \cdot {\pa} \quad , \label{1stk}
\eea
where we have used (\ref{gteq}) to go from (\ref{usegteq}) to the next line,
and the divergenceless condition of $\kappa$ to arrive at the last line.
Notice that $\tilde{\kappa}_{(1)}$ in 
(\ref{1stk})
satisfies the divergenceless condition:
\bea
\partial \cdot \tilde{\kappa}_{(1)}
=
\partial_{i} \partial_{j} 
( \Theta(t) b^{[i} \kappa^{j]} )
= 0.
\eea
Here, $[\, \, \, ]$ denotes the anti-symmetrization in indices.

Next, let us consider $[A,[A,B]]$.
The calculation is almost the same 
to the above, expect that
there is an explicit $t$-dependence in $\tilde{\kappa}_{(1)}(t)$.
We obtain
\bea
 \label{AAB}
[A,[A,B]]
=
\Theta(t) 
\left(
\tilde{\kappa}_{(2)}'(t) + \partial_t \tilde{\kappa}_{(1)}(t)
\right)
\cdot \partial \quad ,
\eea
where 
\bea
\tilde{\kappa}'{}_{(2)}^{\,j}(t) 
\equiv 
\partial_{i}
( \Theta(t) b^{[i} \tilde{\kappa}_{(1)}^{j]}(t) ) .
\eea
However,
the $t$-derivative on $\tilde{\kappa}_{(1)}^i(t)$
does not affect the divergenceless condition
of $\tilde{\kappa}_{(1)}^i(t)$
(as long as $\tilde{\kappa}_{(1)}^i(t)$
is a smooth function of $t$ and $x$).
Thus, (\ref{AAB}) can be again rewritten in a form
\bea
\label{AAB-result}
[A,[A,B]]
=
\Theta(t) 
\left(
\tilde{\kappa}_{(2)}(t)
\right)
\cdot \partial \quad ,
\eea
with divergenceless $\tilde{\kappa}_{(2)}^i(t)$.

Repeating the same arguments, since
the $\Theta(0) = \Theta$ is constant,
it becomes clear that $\hat{\kappa}(t)$ in (\ref{eABk}) satisfies
\bea
 \label{dvlss}
\partial \cdot \hat{\kappa}(t) = 0.
\eea
(\ref{dvlss}) is true for all $t$, including $t=0$.
Thus,
$\hat{\kappa}^i = \hat{\kappa}^i(t=0) $ is divergenceless
when $\kappa$ is divergenceless.

\section{Finite gauge transformations} \label{finiteGT}

Since the gauge symmetry in the commutative side is 
Abelian, 
the
{\em finite}
gauge transformation of $b^{i}$ 
by a finite parameter $\Lambda_i$ 
which we denote as $b^{i}_{\Lambda}$ 
takes
the same form as for the infinitesimal gauge transformations (\ref{gtrbi}):
\begin{equation}
b^{i}_{\Lambda}  =  b^{i} + 
\theta^{ijk} 
\pa_j \Lambda_k .
\end{equation}
Therefore, 
the Seiberg-Witten map of 
the gauge transformed field $b^i_{\Lambda}$ is 
again written in terms of $B$ defined in (\ref{AB})
%
with finite $\Lambda_i$:
\begin{equation}
\hat{b}^{i}_{\hat{\Lambda}} = (e^{A+B} - 1)x^i \Bigr|_{t = 0}.
\end{equation}
This can be rewritten as
\begin{equation}
 \label{bL}
\hat{b}^{i}_{\hat{\Lambda}} = \left(e^{A+B}e^{-A}\hat{b}^{i}(t) 
+ (e^{A+B}e^{-A}-1)x^{i}\right)\Bigr|_{t = 0} ,
\end{equation}
where
\bea
\hat{b}^i (t) \equiv (e^{A} - 1)x^i , \quad \hat{b}^i (t=0) =  \hat{b}^i.
\eea
%
Using the Baker-Campbell-Hausdorf formula, 
(\ref{bL}) can be written as
\begin{equation}
\hat{b}^{i}_{\hat{\Lambda}} = \left(e^{h(A,B)}\hat{b}^{i}(t) 
+ (e^{h(A,B)}-1)x^{i}\right)\Bigr|_{t = 0} ,
\end{equation}
where, as we will show below, $h(A,B)$ takes the form:
\begin{equation}
 \label{formh}
h(A,B) = \Theta(t)\hat{K}^i(t) \pa_i ,
\end{equation}
where $\Theta (t)$ is defined in (\ref{thTht}).
Then, $\hat{b}^i_{\Lambda}$ can be written as
\begin{equation}
 \label{hbt}
\hat{b}^{i}_{\hat{\Lambda}} = 
 \left(e^{\Theta(t)\hat{K}^i(t)\pa_i}  \hat{b}^{i}(t) +
 (e^{\Theta(t)\hat{K}^i(t)\pa_i}-1)x^{i}\right)\Bigr|_{t = 0} .
\end{equation}
As we will show shortly, 
$\hat{K}^i (t)$ 
satisfies the divergenceless condition 
\begin{equation}
\pa_i \hat{K}^i (t) = 0.
\end{equation}
Therefore, 
$\hat{K}^i (t)$ can be written as
\begin{equation}
\hat{K}^i (t) =  \epsilon^{ijk}
\pa_j\hat{\Lambda}'_k (t) .
\end{equation}
Taking $t =0$ in (\ref{hbt}),  
the finite
gauge transformation of the field $\hat{b}^i$ is given as 
\begin{equation}
\hat{b}^{i}_{\hat{\Lambda}} =
\left( e^{\theta^{ljk}(\pa_j\hat{\Lambda}'_k)\pa_l}\hat{b}^{i} 
+\left(
e^{\theta^{ljk}\pa_j\hat{\Lambda}'_k \pa_l} -1 \right) x^{i}\right) ,
\end{equation}
where $\hat{\Lambda}'_k \equiv \hat{\Lambda}'_k (t =0)$, 
and
$\theta^{ijk}$ and $\Theta^{ijk}$
are related as in (\ref{thTh}).
Notice that the reparametrization
from $ \rho(x^i) = x^i + \hat{b}^i$
to $\rho_{\Lambda}(x^i) = x^i + \hat{b}^{i}_{\hat{\Lambda}}$ :
\bea
\rho_{\Lambda}(x^i) =
x^i + \hat{b}^{i}_{\hat{\Lambda}}
=
e^{\theta^{ljk}(\pa_j\hat{\Lambda}'_k)\pa_l}
(x^i + \hat{b}^i)
= e^{\theta^{ljk}(\pa_j\hat{\Lambda}'_k)\pa_l}
\rho(x^i)
,
\eea
is nothing but the finite form of the
volume-preserving diffeomorphism.

\subsection*{The form of $h(A,B)$ in (\ref{formh}) and
the divergenceless condition for $\hat{K}^i$}\label{hK}

From the Baker-Campbell-Hausdorf formula,
$h(A,B)$ is a sum of the terms
which have the form of multiple 
commutations
with $A$ or $B$, for example
\begin{equation}
[\cdots, [B, [A, [A, [B, \cdots[A,B]\cdots]]]]\cdots] \label{Form of multiple}.
\end{equation}
To calculate such terms,
we use the notation in appendix \ref{dlessk}
(see (\ref{thTh}) $\sim$ (\ref{appdefbk})).
According to 
(\ref{1stk}) and 
(\ref{AAB-result}), 
if $Z = \Theta(t) {Z}^{i}(t)\pa_i$
satisfies the divergenceless
condition
$\pa_i Z^i(t) = 0$,
the quantity $[A, Z]$ 
takes the form 
\bea
\label{formAZ}
[A, Z] = \Theta(t) \tilde{Z}^{i}(t) \pa_i,
\eea
with $\pa_i \tilde{Z}(t)^i =0$.
%
Namely,
\begin{equation}
\pa_i Z^i(t) = 0 \Rightarrow \pa_i \tilde{Z}^i(t) = 0 . \label{div A}
\end{equation}
%
Next, we show that 
if $Z = \Theta(t)Z^i(t) \pa_i$ satisfies the divergenceless condition $\pa_i Z^{i}(t)=0$,
$[B, Z]$ also takes the form 
\bea
[B, Z]= \Theta(t) \tilde{Z}'^{i}(t)\pa_i, \label{formBZ}
\eea
with $\pa_i \tilde{Z}'^{i}(t)=0$.
Namely,
\begin{equation}
\pa_i  {Z}^{i}(t) = 0 \Rightarrow \pa_i  \tilde{Z}'^{i}(t) = 0. \label{div B}
\end{equation}
By a direct calculation, we have
\begin{eqnarray}
 \label{BZ}
[B,Z] &=& [\Theta(t) \kappa^j \pa_j, \Theta(t)Z^i(t) \pa_i ] \nn \\
&=& \Theta(t)\left(\kappa^j \pa_j\Theta(t)Z^i(t)
+\Theta(t)\kappa^j \pa_j Z^i(t)
\right)\pa_i -\left(\kappa 
 \leftrightarrow Z (t) \right) \nn \\
&\equiv&
\Theta (t) \tilde{Z}'^{i}(t) \pa_i .
\end{eqnarray}
Here, we used $B$ defined in (\ref{gAB}), and
$\kappa^i$ is the one defined in (\ref{appdefbk})
which is a different notation from the main body.
From (\ref{BZ}) we see that $[B,Z]$ has the form (\ref{formBZ}).
On the other hand, the divergence of $\tilde{Z}'^{i}(t)$ becomes
\begin{eqnarray}
\pa_i \tilde{Z}'^{i}(t) &=& 
\pa_i\kappa^j \pa_j\Theta(t)Z^i(t)  
+  \kappa^j Z^i(t) \pa_j\pa_i \Theta(t) 
+ \pa_i \Theta(t) \kappa^j \pa_j Z^i(t)
+\Theta(t)\pa_i \kappa^j \pa_j Z^i(t)
\nn  \\
&& -(\kappa 
\leftrightarrow Z(t)).
\label{div Z}
\end{eqnarray}
Since the first line of (\ref{div Z}) is symmetric under the exchange 
$\kappa^i  
\leftrightarrow Z^i(t)$,
the right hand side of (\ref{div Z}) vanishes:
\begin{equation}
\pa_i \tilde{Z}'^{i}(t) = 0. \label{divergenceless proof}
\end{equation}
Thus we 
have proved 
(\ref{div B}).
From (\ref{formAZ}) and (\ref{formBZ}),
$h(A,B)$ can be written in a form in (\ref{formh}):
\bea
h(A,B) = \Theta(t)\hat{K}^i(t) \pa_i ,
\eea
where from (\ref{div A}) and (\ref{div B})
$\hat{K}^i(t)$ is divergenceless:
\begin{equation}
\pa_i \hat{K}^i (t) = 0,
\end{equation}
which also holds for $t=0$.


\section{Divergenceless condition for $\hat{B}_{\mu}{}^i$}\label{dlessB}


We first introduce a short hand notation
for the adjoint action of ${A}$ 
which we will use repeatedly:
\begin{equation}
\A C \equiv [{A}, C] ,
\end{equation}
for any operator $C$.


Let us consider the operator appearing in (\ref{allordB}):
\begin{equation}
 \label{allordB2}
\hat{B}_{\mu}{}^{{i}} \pa_{{i}} 
\equiv  \left(e^{{A}} (\hpa_{\mu} - \Theta(t)(B_{\mu}{}^{{i}} - \pa_{\mu}b^{{i}}
 )\hpa_{{i}} )
e^{-{A}} - \hpa_{\mu}\right)\Bigr|_{t = 0},
\end{equation}
where $B_{\mu}{}^{i} \equiv \e^{ijk} \pa_j b_{\mu k}$.
As in the appendix \ref{dlessk},
in this appendix we use notation for $B_{\mu}{}^i$
different from that in the main body (\ref{BhB})
by the overall scaling.
We can decompose the operator 
in the right hand side of (\ref{allordB2}) (before taking $t=0$)
into three parts:
\begin{enumerate}
 \item[I.] The term arising from the exponential map of $\hpa_{\mu}$.
 \item[II.] The term arising from the exponential map of  $\Theta(t) B_{\mu}{}^{{i}}$.
 \item[III.] The term arising from the exponential map of $\Theta(t) \pa_{\mu}b^{{i}}$.
\end{enumerate}
As discussed in the previous section,
the term [II] satisfies the divergenceless condition by itself
since $\pa_{i} B_{\mu}{}^i = 0$.
Therefore, the remaining thing to show is that
[I]+[III] satisfies the divergenceless condition.

Let us first look at [I] 
(here we also include the term $-\pa_\mu$ in the right hand side
of (\ref{allordB2}) in this category):
\begin{eqnarray}
e^{{A}} \hpa_{\mu} e^{-{A}}  - \hpa_{\mu} &=& 
\sum_{n = 0}^{\infty} 
\frac{1}{n!} \A^{n}\hpa_{\mu} - \hpa_{\mu} \nn \\
&=& \sum_{n = 1}^{\infty}\frac{1}{n!}\A^{n-1}(-\pa_{\mu}(\Theta(t)
b^{{i}}\pa_{{i}})) \nn \\
&=& \sum_{n = 0}^{\infty}  
\frac{1}{(n+1)!}
\A^{n}\left(-t\Theta(t) (\Theta(t)\pa_{\mu}(\pa \cdot b)
b^{{i}}\pa_{{i}})
- \Theta(t)(\pa_{\mu}b^{{i}}\pa_{{i}})\right)\label{partial mu} ,
\end{eqnarray}
where we have used
\begin{equation}
\pa_{\mu}\Theta(t) = t \Theta^2(t)\pa_{\mu}(\pa_{{i}}b^{{i}}),
\end{equation}
which follows directly from (\ref{gteq}).
As a short hand notation, we define
\begin{equation}
 \label{defalpha}
\alpha_{\mu} \equiv (\Theta(t)\pa_{\mu}(\pa \cdot b)b^{{i}}\pa_{{i}})
\equiv \alpha_{\mu}^{{i}}\pa_{{i}},
\end{equation}
\begin{equation}
 \label{defbeta}
\beta_{\mu} \equiv (\pa_{\mu}b^{{i}}\pa_{{i}})
\equiv \beta_{\mu}^{{i}}\pa_{{i}}.
\end{equation}
Then, (\ref{partial mu}) can be written as
\begin{eqnarray}
&&\sum_{n = 0}^{\infty}  
\frac{1}{(n+1)!}
\A^{n}\left(-t\Theta(t) (\Theta(t)\pa_{\mu}(\pa \cdot b)
b^{{i}}\pa_{{i}})
- \Theta(t)(\pa_{\mu}b^{{i}}\pa_{{i}})\right)
\nn \\ 
&=& \sum_{n = 0}^{\infty}  
\frac{1}{(n+1)!}
\A^{n}\left(-t\Theta(t) \alpha_{\mu}
- \Theta(t)\beta_{\mu}\right)
\nn \\
&=&   -t \Theta(t)\alpha_{\mu} - \Theta(t)\beta_{\mu} \nn\\
&& + \sum_{n = 1}^{\infty}   
\frac{1}{(n+1)!}
\left(-t\A^{n}(\Theta(t) \alpha_{\mu})
-n \A^{n-1}(\Theta(t) \alpha_{\mu})
- \A^{n}(\Theta(t)\beta_{\mu})\right).
\nn \\
\end{eqnarray}
When we take $t = 0$,
the relevant part of the operator is given by
\begin{equation}
e^{{A}} \hpa_{\mu} e^{-{A}}  - \hpa_{\mu} 
= - \Theta(t)\beta_{\mu}
+ \sum_{n = 1}^{\infty}   
\frac{1}{(n+1)!}
\left(-n \A^{n-1}(\Theta(t) \alpha_{\mu})
- \A^{n}(\Theta(t)\beta_{\mu})\right) . \label{part-I}
\end{equation}

%

Next, we turn to the term [III] which is given by
\begin{equation}
e^{{A}}(\Theta(t) \pa_{\mu}b^{{i}}\hpa_{{i}}) e^{-{A}} 
= \Theta(t)\pa_{\mu}b^{{i}}\pa_{{i}} +
\sum_{n = 1}^{\infty}   
\frac{1}{n!}\A^n
(\Theta(t)\pa_{\mu} b^{{i}}\pa_{{i}}) .
\end{equation}
Using $\beta_{\mu}$ in (\ref{defbeta}), it is written as
\begin{equation}
e^{{A}}(\Theta(t) \pa_{\mu}b^{{i}}\hpa_{{i}}) e^{-{A}} 
=  \Theta(t)\beta_{\mu}
+ \sum_{n = 1}^{\infty}   
\frac{1}{n!}\A^n(\Theta(t)\beta_{\mu}) . \label{part-III}
\end{equation}

Now we show that
the sum of (\ref{part-I}) and (\ref{part-III}) is divergenceless.
It 
is given by
\begin{equation}
+ \sum_{n = 1}^{\infty} 
\left(\frac{1}{n!}\A^n(\Theta(t)\beta_{\mu}) 
-\frac{1}{(n+1)!}
\left(n \A^{n-1}(\Theta(t) \alpha_{\mu})
+\A^{n}(\Theta(t)\beta_{\mu})\right)\right) ,
\end{equation}
which becomes
\begin{equation}
- \sum_{n = 1}^{\infty} 
\left(
\frac{n}{(n+1)!}
\A^{n-1}\left((\Theta(t) \alpha_{\mu})
-\A(\Theta(t)\beta_{\mu})\right)\right) .
\end{equation}
Thus we need to show that $\gamma_{\mu}^{{i}}$ defined by
\begin{equation}
\Theta(t)\gamma_{\mu}^{{i}}\pa_{{i}} \equiv (\Theta(t) \alpha_{\mu})
-\A(\Theta(t)\beta_{\mu}) ,
\end{equation}
satisfies the 
divergenceless condition:
\begin{equation}
\pa_{{i}}\gamma_{\mu}^{{i}} = 0. \label{divergenceless}
\end{equation}
This can be shown by a direct calculation.

\bibliography{M5SWmap}
\bibliographystyle{utphys}

\end{document}